\documentclass[twocolumn,floatfix,prb,aps,showpacs]{revtex4}

\usepackage{graphicx}
\usepackage{amsfonts}
\usepackage{amsmath}
\usepackage{multirow}
\usepackage{pbox}
\usepackage{epstopdf}
\usepackage{xfrac}

\begin{document}

\title{Emergence of Jack ground states from two-body pseudopotentials in fractional quantum Hall systems}

\author{Bartosz Ku\'smierz and Arkadiusz W\'ojs}

\address{Department of Theoretical Physics,Wroc{\l}aw University of Science and Technology, Wybrze\.ze Wyspia\'nskiego 27, 50-370 Wroc{\l}aw, Poland}

\date{\today}

\begin{abstract}
The family of ``Jack states'' related to antisymmetric Jack polynomials are the exact zero-energy ground states of particular model short-range {\em many-body} repulsive interactions, defined by a few non-vanishing leading pseudopotentials.
Some Jack states are known or anticipated to accurately describe many-electron incompressible ground states emergent from the {\em two-body} Coulomb repulsion in fractional quantum Hall effect.
By extensive numerical diagonalization we demonstrate emergence of Jack states from suitable pair interactions.
We find empirically a simple formula for the optimal two-body pseudopotentials for the series of most prominent Jack states generated by {\em contact} many-body repulsion.
Furthermore, we seek realization of arbitrary Jack states in realistic quantum Hall systems with Coulomb interaction, i.e., in partially filled lowest and excited Landau levels in quasi-two-dimensional layers of conventional semiconductors like GaAs or in graphene.
\end{abstract}


\maketitle

\section{Introduction}
\label{sec1}

Over three decades after its discovery\cite{Tsui82}, the fractional quantum Hall effect (FQHE) remains one of the most intriguing phenomena in condensed matter physics. 
Present understanding\cite{DasSarma97,Yoshioka02,Jain07} of this remarkable collective behavior of strongly correlated quasi-two-dimensional (2D) electrons in high magnetic field has involved many new concepts, most importantly that of Jain composite fermions (CFs)\cite{Jain89}, i.e., bound states of electrons and vortices of the many-electron wave function, weakly interacting through residual forces and filling effective Landau levels (LL) of greatly reduced degeneracy.
An important direction in FQHE studies have always been attempts to find model wave functions describing incompressible many-electron phases realized in real experimental conditions.
Famous examples are the Laughlin\cite{Laughlin83}, Moore-Read\cite{Moore91} (Pfaffian) and Read-Rezayi\cite{Read99} (parafermion) wave functions corresponding to the particular Landau level (LL) filling factors $\nu=1/2$, $1/2$, and $3/5$, respectively.
While all these wave functions can be elegantly understood in terms of either non-interacting or simply correlated CFs\cite{Jain07,Scarola00,Sreejith11}, the original ideas often came from somewhere else.

All model FQHE wave functions describe a partially filled (lowest or higher) LL, in some cases folded with respect to spin or multiple (iso)spins, but in this work we will assume full LL polarization and ignore this additional degeneracy.
As a partially filled higher LL can be mapped onto the lowest LL (with the same filling factor $\nu$ and the same pseudopotential\cite{Haldane83} expressing pair interaction energy $V$ as a function of relative pair angular momentum $m$), the wave functions are often defined in the latter.
And as the single-electron orbitals of the lowest LL are (in symetric gauge) simply the monomials in the complex coordinate $z=x+iy$ indexed by angular momentum, $\phi_m(z)\sim z^m$, the relevant many-electron wave functions are sought in the form of anti-symmetric complex polynomials (of an infinite number of variables $z_i$ and an infinite degree, connected through a finite $\nu$).

A broad class of FQHE wave functions called ``Jack states'' has been derived from the theory of symmetric polynomials\cite{Macdonald95,Macdonald88,Stanley88,Kerov03}.
The above-mentioned Laughlin, Pfaffian, and parafermion states are all members of the Jack family, corresponding to rather simple root occupations [100], [1100], and [11100], respectively, and their identification as such provided new insight\cite{Bernevig08a,Bernevig08b,Bernevig09a,Bernevig09b,Thomale11} and an explicit construction method based on the recursion relations between the Jack expansion coefficients in the relevant (Slater determinant) basis\cite{Sogo94,Lapointe00}.

A useful property of the above three and some other Jack states important in the context of FQHE is that they are exact zero-energy ground states of certain model many-body interactions.
A notion of a pair pseudopotential can be extended\cite{Wojs05,Simon07a,Simon07b} to a many-body interaction in an isolated LL: the $K$-body pseudopotential $V(m)$ is the $K$-body interaction energy $V$ as a function of $K$-body relative angular momentum $m$.
In this language, the Laughlin, Pfaffian, and parafermion wave functions are generated by a two-, three-, and four-body contact repulsion correspondig to pseudopotentials with only one non-zero (positive) leading coefficient, and other wave functions are generated analogously by more complex many-body pseudopotentials\cite{Moore91,Read99,Wojs05,Simon07a,Simon07b,Trugman85}.

In this paper we examine two questions: 
(i) Can Jack states, which are generated exactly by particular short-range {\em many-body} repulsion, emerge also as approximate ground states of suitable {\em two-body} repulsion? 
(ii) Do various Jack states describe Coulomb ground states in different LLs (in conventional semiconductors or in graphene), and thus are relevant description of the incompressible quantum liquids of FQHE?

We perform extensive numerical calculations by means of exact diagonalization in Haldane spherical geometry\cite{Haldane83} to obtain the quasi-continua of ground states of arbitrary short-range two-body pseudopotentials $V(m)$ for many relevant finite systems of $N$ electrons at magnetic flux $2Q$.
Then we use the theory of Jack polynomials to semi-analytically construct the Jack states on the plane\cite{Sogo94,Lapointe00}, and then through stereographic projection\cite{Fano86} transform them into spherical geometry. 
In some cases we also employ many-body (for up to $K=5$) exact diagonalization\cite{Wojs05} to compute the Jack states directly on a sphere.
Finally, we compare the Jack states with the maps of two-body ground states and with the Coulomb ground states, by studying the overlaps and pair-correlation functions.
Indirect comparison of Jack and Coulomb ground states through the maps of overlaps with ground states of arbitrary pair interaction allows a more secure conclusion about their connection, especially when the direct overlap is not convincingly high or it is sensitive to small variation of the Coulomb interaction.

The main result answers the above question (i): 
We demonstrate that Jack exact ground states of short-range $K$-body repulsions are in general accurately reproduced by the suitable short-range two-body interaction. 
In particular, we find a simple formula for the pair pseudopotentials mimicking the many-body contact repulsion, linking the range $m$ of the former with the order $K$ of the latter.
Furthermore, regarding question (ii), we show that ground states of the long-range Coulomb pseudopotential are represented with excellent accuracy by a suitable short-range model, but only few (already known) Jack states can emerge in realistic Coulomb systems in GaAs or (monolayer) graphene.

The paper is organized as follows: 
In the next Section \ref{sec2} we briefly overview Jack polynomials and standard tools used in the symmetric function theory, and discuss Jack states in the context of many-body interactions. 
Main results are presented in Section \ref{sec3} in form of the series of tables and maps of overlaps. 
These data are then used to indicate what pair pseudopotentials generate Jack states, and what Jack states are viable trial functions for the FQHE. 
In the last Section \ref{sec4} we conclude our studies. 

\section{Jack states}
\label{sec2}

The Jack polynomial\cite{Macdonald95,Macdonald88,Stanley88,Kerov03,Sogo94,Lapointe00,Hora07,Baratta11,Kusmierz14,Kusmierz16a,Kusmierz16b,Kusmierz16c,Gioacchino17}, called simply a ``Jack'' and denoted by $J^{\alpha}_{\lambda}$, is a symmetric polynomial indexed by the partition $\lambda$ and the real number $\alpha$. 
The {\em partition} is a sequence $\lambda=(\lambda_1, \lambda_2,\dots)$ of non-negative integers in the non-increasing order. 
The non-zero elements of the sequence are called {\em parts} of partition $\lambda$. 
The number of parts is the {\em length} of partition $\lambda$ and it is denoted by $\ell(\lambda)$. 
The symbol $m(\lambda,i)$ means the number of parts of partition $\lambda$ equal to $i$. 
One defines the {\em natural order} on partitions and says that $\lambda$ {\em dominates} $\mu$ if for every natural number $i$ the sum of the first $i$ parts of $\lambda$ is greater or equal then the sum of its first parts of $\mu$. 
Such relation is denoted as $\lambda \geq \mu$. 
The addition of two partitions is defined by adding parts indexed by the same numbers $(\lambda + \mu)_i = \lambda_i + \mu_i$. 
In the context of FQHE it is useful to represent partitions in the occupation-number configuration $\lambda = [m(\lambda, 0) \; m(\lambda, 1)\; \dots]$.

Monomial symmetric functions, i.e., ``monomials'' $m_{\lambda} (z_1, z_2, \dots, z_N) \equiv m_\lambda$ are defined as
\begin{eqnarray}
m_\lambda &=& \left( m(\lambda ,0)! \cdot m(\lambda ,1)! \cdot  \dots \cdot m(\lambda ,N)! \right)^{-1} 
\nonumber
\\ 
&\times& \sum_{\sigma \in S_N} {z}_1^{\lambda_{\sigma(1)}} {z}_2^{\lambda_{\sigma(2)}} \dots {z}_N^{\lambda_{\sigma(N)}}. 
\end{eqnarray}
The monomials are standard basis in the ring of symmetric functions. 
Jacks $J^{\alpha}_{\lambda}$ can be defined as eigenfunctions of the differential Laplace-Beltrami operator $H_{\rm LB}$ indexed by a real number $\alpha$
\begin{equation}
H_{\rm LB}({\alpha}) = \sum_{i=1}^N (z_i\partial_i)^2 + \frac{1}{\alpha} \sum_{1\leq i<j\leq N} 
\frac{z_i+z_j}{z_i-z_j} (z_i\partial_i-z_j\partial_j).
\end{equation}
Its eigenvalues are given by: 
\begin{equation}
E_{\lambda} = \sum_{i=1}^{\ell(\lambda)} \left( \lambda_i^2 + \frac{1}{\alpha}(N+1-2i) \lambda_i \right).
\end{equation}
When expanded in the monomial basis, Jacks reveal non-zero coefficients only for the monomials indexed by partitions dominated by the Jack's root partition: $J_{\lambda}^{\alpha} = \sum_{\mu \leq \lambda} m_{\mu} u_{\lambda \mu}(\alpha)$ ($v_{\lambda \mu}(\alpha) \in \mathbb{R}$). 
Under the normalization condition $v_{\lambda \lambda}=1$, coefficients $v_{\lambda \mu}$ are the inverse polynomials in $\alpha$ and have no roots for $\alpha > 0$. 
Furthermore, for a fixed partition $\lambda$, the Jack $J_{\lambda}^{\alpha}$ is well-defined for all but a finite number of negative values of $\alpha$. 
The points at which the Jack is undefined are called poles. 
The recursion formula for the coefficients of a Jack in the monomial basis has been derived\cite{Sogo94,Lapointe00}.

Jack fermionic polynomials\cite{Bernevig09b,Thomale11} $S_{\mu}^{\alpha}$ are antisymmetric analogues of Jack symmetric polynomials. 
They are defined as a product of a symmetric Jack and the Vandermonde determinant (multiplication by the Vandermonde determinant is a canonical isomorphism of the ring of symmetric polynomials on the ring of antisymmetric polynomials)
\begin{equation}
S_{\lambda+\delta}^{\alpha} (z_1,\dots,z_N) = J_{\lambda}^{\alpha}(z_1,\dots,z_N) \prod_{i < j}^{N} (z_i - z_j),
\end{equation}
where $\delta = (N-1,N-2,\dots,1,0)$. 
Fermionic Jacks are eigenvectors of the fermionic Laplace-Beltrami operator
\begin{equation}
H^{\rm F}_{\rm LB} (\alpha)= H_{\rm KIN} + \left ( \frac{1}{\alpha} - 1 \right ) H_{\rm INT},
\end{equation}
where
\begin{equation}
H_{\rm KIN} = \sum_{i=1}^N (z_i\partial_i)(z_i\partial_i),
\end{equation}
and
\begin{equation}
H_{\rm INT} = \sum_{1 \leq i<j \leq N} \frac{z_i+z_j}{z_i-z_j}(z_i\partial_i-z_j\partial_j)-2\frac{z_i^2+z_j^2}{(z_i-z_j)^2}.
\end{equation}
Recursion formula for fermionic Jacks in terms of Slater determinants has been derived\cite{Bernevig09b,Thomale11}.

Standard basis in the ring of antisymmetric polynomials are Slater determinants $\text{sl}_{\mu}$
\begin{equation}
\text{sl}_{\mu} (z_1,z_2,\dots,z_N) = \sum_{\sigma \in S_n} \text{sgn}(\sigma) \cdot z_{\sigma(1)}^{\mu_1} \cdot z_{\sigma(2)}^{\mu_2} \cdot \dots \cdot z_{\sigma(N)}^{\mu_N}.
\end{equation}

Jack states are FQH states related to the Jack polynomials. 
As it has been pointed out earlier\cite{Bernevig08a,Bernevig08b,Bernevig09a}, the analysis of angular momentum operator imposes certain necessary condition on both partition and real parameter of valid Jack states. 
Bernevig and Haldane\cite{Bernevig08a,Bernevig08b,Bernevig09a} considered condition of uniformity on the sphere (highest weight and lowest weight) for bosonic wave functions and established what follows. 
The real parameter $\alpha_{k,r} = -(k + 1)/(r - 1)$ for $(k + 1)$ and $(r - 1)$ both positive, and coprime, the partition length equals $ \lambda_{\ell(\lambda)} = (r - 1)s + 1$ and the partition itself is of the form $\lambda = [n_0, 0^{s(r-1)},k,0^{r-1},k, 0^{r-1},k,\dots,k ]$. 
Here, $0^{r-1}$ means a sequence of $r-1$ zeros and $n_0$ is certain natural number. 
Such partition is denoted as $\lambda^0_{k,r,s}$. 
The case of $s=1$ provides many FQH ground states and the cases $s>1$ are related to quasiparticle states. 
In this paper we focus on the ground states. 
We denote the partitions by $\lambda^0_{k,r,s=1} = \lambda^0_{k,r}$. 
The Jacks indexed by $\lambda^0_{k,r}$ and $\alpha_{k,r}$ are related to boson FQH states at filling factor $\nu = k/r$.

For example the bosonic Laughlin wave function for the state $\nu = 1/q$ ($q$ even) can be represented as a product of the Gaussian and the following symmetric Jack:
\begin{equation}
\Psi_{\rm L}^q = \prod_{i < j}^{N} (z_i - z_j)^q = J_{\lambda^0(1,r)}^{\alpha_{1,q}} .
\end{equation}
As it trivially follows, fermionic Laughlin wave functions for state $1/q$ also are Jack states for partition $\lambda^0(1,q)$ and real parameter $\alpha_{1,q-1}$. 
Laughlin wave function can be described in terms of non-interacting composite fermions (see the following subsection).

The Moore-Read (Pfaffian) state, which for bosons occurs at $\nu=1$ and for fermions at $\nu=1/2$, and reads:
\begin{equation}
\Psi^m_{\rm MR} = {\text{Pf}} \left ( \frac{1}{z_i-z_j} \right ) \prod_{i < j}^{N} (z_i - z_j)^{m+1},
\end{equation}
is well-defined for even numbers of particles and can be written as either bosonic or fermionic Jack:
\begin{equation}
\Psi^0_{\rm MR} = J^{\alpha_{2,2}}_{\lambda^{0}_{(2,2)}} \;\;\;{\rm or} \;\;\; \Psi^1_{\rm MR} = S^{\alpha_{2,2}}_{\lambda^{0}_{(2,2)}+\delta}.
\end{equation}
The other Jack states include the Read-Rezayi (parafermion) state and the Gaffnian. 

\section{Comparison with two-body ground states}
\label{sec3}

Let us now turn to resolving two principal questions of this research announced already in the Introduction:

(Q1) Can Jack states, which are generated by particular short-range {\em multi-particle} repulsion, emerge also as ground states of suitable {\em two-body} Hamiltonians? 

(Q2) In particular, do various Jack states describe Coulomb ground states in different LLs (in conventional semiconductors or in graphene), and thus are relevant description of the incompressible quantum liquids of FQHE?

\subsection{Exact diagonalization in spherical geometry}

We have explored these questions by a systematic numerical search of suitable two-body Hamiltonians.
For all computations we used standard Haldane spherical geometry\cite{Haldane83,Fano86}, in which $N$ electrons are confined to the surface of a sphere of radius $R$, with radial magnetic field $B$ being generated by a Dirac magnetic monopole of strength $2Q\,hc/e$, corresponding to the magnetic length $l_B=R/\sqrt{Q}$.
In this geometry, consecitive LLs denoted as LL$_n$ appear in the form of single-particle angular momentum shells (lengths $l=Q+n$, $n=0$, 1, \dots; projections $|m|\le l$).
In particular, the lowest LL with $n=0$, denoted as LL$_0$, corresponds to angular momentum $l=Q$ and has degeneracy of $2Q+1$.

The $N$-electron Hilbert space is spanned by the configurations $\left|m_1,m_2,\dots,m_N\right>$, and the two-body interaction matrix elements are connected to two-body Haldane pseudopotential\cite{Haldane83} $V(m)\equiv V_m$, which defines pair interaction energy $V$ as a function of relative pair angular momentum $m=1$, 3, 5, \dots, through the Clebsch-Gordan coefficients: $\left<m_1',m_2'|V|m_1,m_2\right>=\sum_m\left<m_1',m_2'|L\right>V_m\left<L|V|m_1,m_2\right>$, where $L=2l-m$ is the total pair angular momentum on a sphere.

Hamiltonians defined by interaction $V_m$ are diagonalized numerically with simultaneous resolution of total angular momentum $L$ using a variant of the nested Lanczos algorithm (resolving $L$ is important, as only the $L=0$ ground states have uniform charge distribution and hence are the possible candidates for the non-degenerate ground states of FQHE).
This is essentially the configuration-interaction method, with the efficiency crucially dependent on the fast implementation of the action of the Hamiltonian on a trial state vector.
(Our codes and today's workstations allow diagonalization of two-body Hamiltonians with dimensions up to several billion.)

\subsection{Model Hamiltonians}

The main calculation consisted of comparing a particular Jack state with the map of computed ground states of fairly arbitrary pair Hamiltonians $H$.
Since in the end we aim to find connection of Jacks with the Coulomb ground states of FQHE, and since the latter are known to be essentially determined by the short-range part of the relevant Coulomb pseudopotential, we restrict our search of suitable pair Hamiltonians to the model pseudopotentials which vanish for $m>5$ (except for the case of $\nu=1/5$ as explained in Sec.~\ref{Jack10000}).
With the overall scale being irrelevant, we can use an obvious normalization $V_1+V_3+V_5=1$, leaving only two independent parameters of the model and allowing convenient graphical representation of the results.
So the main results will be plotted in form of triangular maps, where each point corresponds to particular ratios between $V_1$, $V_3$, and $V_5$, with all higher pseudopotential coefficients vanishing.
It is quite obvious that with a suitable choice of $V_1:V_3:V_5$ this model will accurately reproduce Coulomb ground states of FQHE; here we are asking whether it can also reproduce the Jack ground states of multi-particle repulsion.

For comparison with Coulomb ground states, we have used Haldane pseudopotentials $V_m\equiv\left<L|V|L\right>$ of the Coulomb interaction potential $V(r)=1/r$, which are calculated separately for each considered LL, in GaAs or graphene.
In GaAs we also consider finite layer width $w$, included by assumming an infinite square-well potential, i.e., the density profile of the form $\varrho(z)\propto\cos^2\pi z/w$.
There is a certain complication with defining a pseudopotential for excited LLs in graphene on a sphere; here we have used the definition of Ref.~\onlinecite{Wojs11}.
Following standard convention, when considering higher LLs (in GaAs or graphene) we map them onto the lowest LL with $l=Q$, retaining the correct pseudopotentials $V_m$ of the given (i.e., excited) LL.

\subsection{Triangular maps}

\begin{figure}
\includegraphics[width=0.40\textwidth]{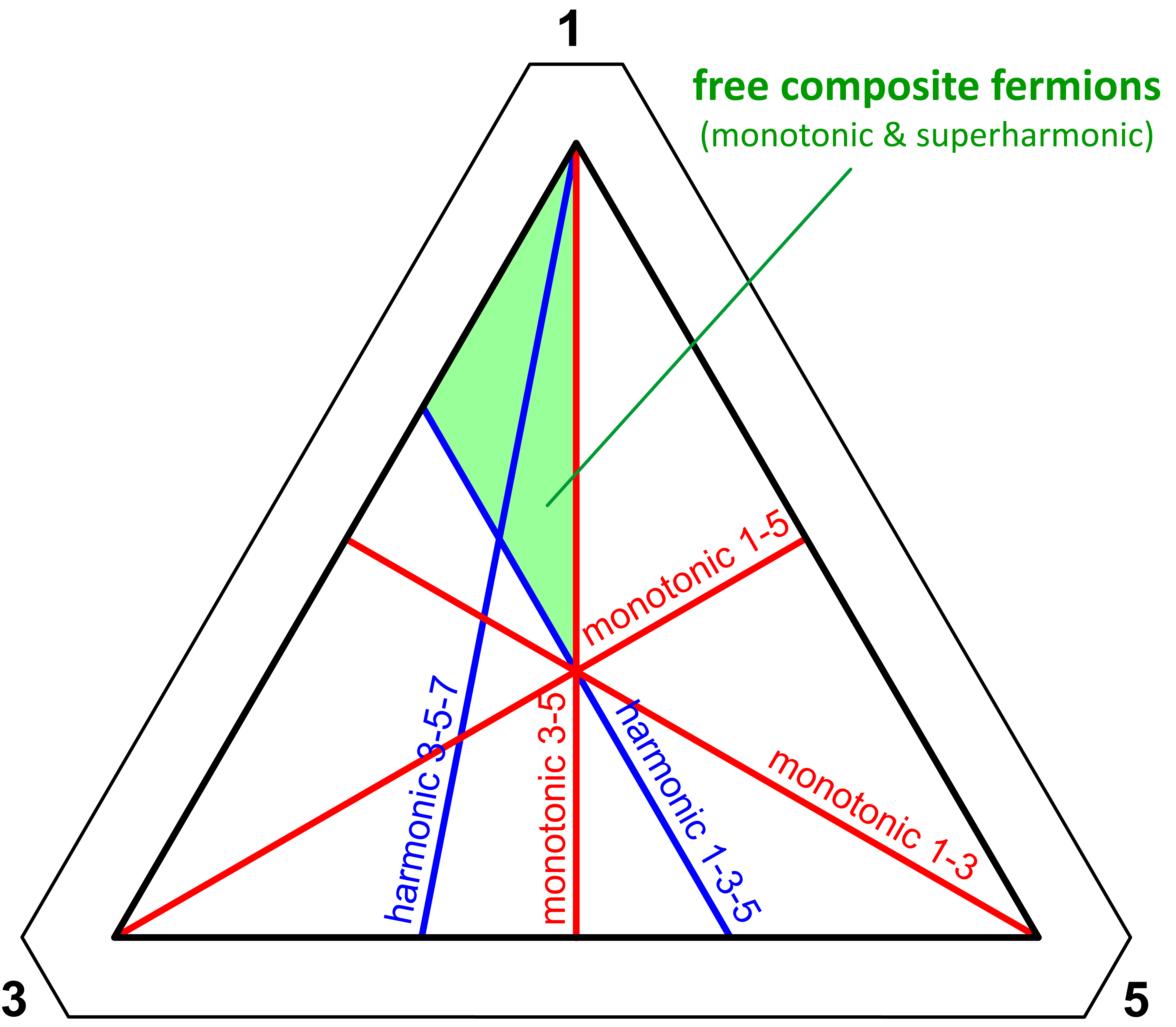}
\caption{
Empty triangular map. 
Each point on the map corresponds to a particular model interaction defined by the values of three pseudopotentials $(V_1,V_3,V_5)$ determined by the distance from three sides of the triangle.
The lines define the points of the triangle in which the pseudopotential is monotonic (or harmonic) through the two (or three) indicated values of $m$.
The green part of the triangle represents the family of pseudopotentials which are both decreasing and superharmonic throughout the short range.
On maps like this in the following figures we will present overlaps of the $L=0$ ground state of the continuously varied pseudopotential $(V_1,V_3,V_5)$ denoted as $\psi(V_1,V_3,V_5)$ with the particular states of interest, such as the Jacks or Coulomb ground states.}
\label{fig1}
\end{figure}

Let us begin with becoming familiar with the triangular map used for the presentation of our main results.
An ``empty'' map is shown in Fig.~\ref{fig1}.

Each corner of the inner triangle corresponds to one positive pseudopotential coefficient $V_m=1$, $m=1$, 3, or 5, as indicated, and all others vanishing.
In other words, the three corners are $(V_1,V_3,V_5)=(1,0,0)$, $(0,1,0)$, and $(0,0,1)$.
Further, the edges of the triangle have two positive coefficients:
$(V_1,V_3,V_5)=(x,1-x,0)$, $(0,x,1-x)$, and $(x,0,1-x)$, with $0<x<1$;
the interior of the triangle has all three coefficients positive (each one proportional to the distance from the respective edge), with the central point obviously corresponding to $(V_1,V_3,V_5)=(1/3,1/3,1/3)$;
and the outside of the triangle represents pseudopotentials with at least one negative coefficient.

For each point on the map, i.e., for each pseudopotential $(V_1,V_3,V_5)$, we have calculated the lowest state at $L=0$ (i.e., uniform) of various systems $(N,2Q)$.
This state is denoted by $\psi_{N,2Q}^{L=0}(V_1,V_3,V_5)$ or shortly $\psi(V_1,V_3,V_5)$.
Thus, the triangular map is not only the map of pseudopotentials, but also the map of the corresponding states $\psi$ (i.e., of the types of correlation) and in that map in the following figures we will display the overlaps of $\psi$ with with the particular states of interest, such as the Jack or Coulomb ground states, for a specific finite-size system $(N,2Q)$.

Correlations in a degenerate LL (and thus in particular the emergence of a particular incompressible ground state) mostly depend onthe monotonocity and harmonicity\cite{Quinn00} of the pseudopotential over the range where it is strong (i.e., usually, for small $m$).
The monotonicity conditions are obvious; the red lines in Fig.~\ref{fig1}, labeled as ``monotonic A-B'' and corresponding to $V_A=V_B$ identify the areas on the map with all possible orderings of $V_1$, $V_3$, and $V_5$.
The ``superharmonicity'' through a series of three $m=A<B<C$ simply means superlinear (convex) dependence over this range, i.e.:
$(V_A-V_B)/(B-A)>(V_B-V_C)/(C-B)$.
The name reflects the fact that a pseudopotential $V_m$ which is linear in $m$ corresponds to a potential $V(r)$ which is linear in $r^2$ (i.e., ``harmonic'') in any LL.
In Fig.~\ref{fig1}, the blue lines labeled as ``harmonic A-B-C'' define the areas on the map with respect to superharmonicity through $m=1$, 3, 5, and through $m=3$, 5, 7 (recall that $V_7\equiv0$).
Importantly, the pseudopotential must be both monotonic and superharmonic at short range (like, e.g., Coulomb interaction in the lowest LL) to support the Laughlin state of essentially free composite fermions at $\nu=1/3$, while the harmonic behavior through $m=1$, 3, and 5 (like, e.g., Coulomb interaction in the second LL in GaAs) results in composite fermion pairing and stabilizes the Pfaffian ground state at $\nu=1/2$.
Thus, it helps to keep in mind the arrangement of red and blue lines on the map when relating the short-range model $(V_1,V_3,V_5)$ with the actual Coulomb pseudopotentials.

\subsection{Results for Jack states generated by two-body repulsion}

The main numerical results, regarding search of Jack states in two-body (especially Coulomb) Hamiltonians, are presented in the following sequence of maps.
To identify different Jacks, we adopted an abbreviated and $N$-independent notation for the root occupations, in which $[100]\equiv[(100)_{N-1}1]$, $[11000]\equiv[(11000)_{(N-2)/2}11]$, etc., i.e., the sequence given in square brackets $[\dots]$ is meant to be repeated so many times as to give correct $N$ and then appended so as to restore the reflection symmetry.

\subsubsection{Jack state $[100]$ (Laughlin 1/3)}

\begin{figure}
\includegraphics[width=0.40\textwidth]{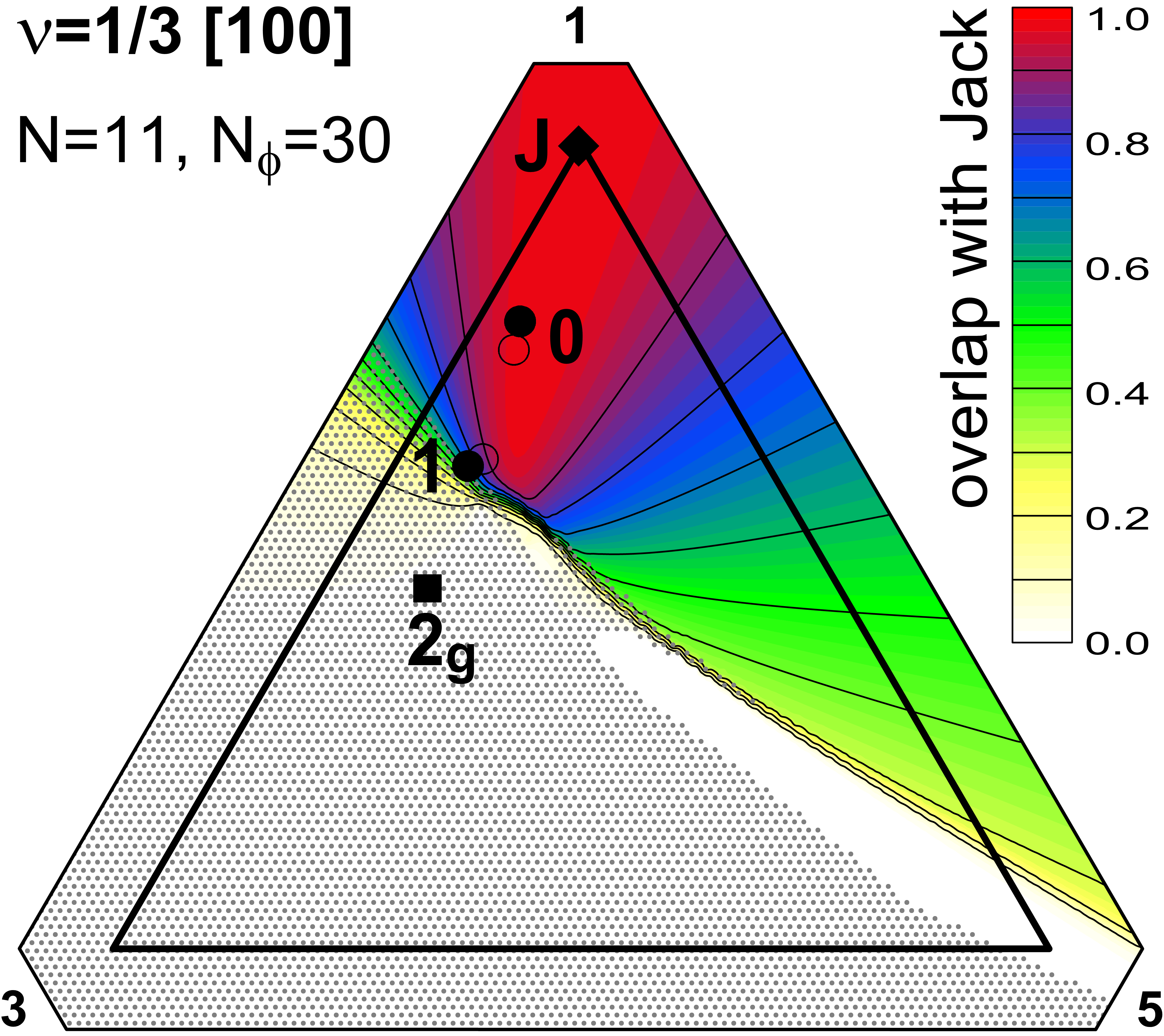}
\caption{
Map of the overlap of Jack state [100] (i.e., the Laughlin $\nu=1/3$ state, generated by two-body repusion at $m=1$) with the lowest $L=0$ states of all possible model short-range two-body pseudopotentials, for the system with $N=11$ and $2Q=30$.
Each point on the map corresponds to a particular pseudopotential and its lowest $L=0$ eigenstate $\psi(V_1,V_3,V_5)$, as explained in Fig.~\ref{fig1}. 
The color at this point indicates the overlap of $\psi$ with the Jack state [100].
Grey dots mark the area on the map in which the absolute ground state has $L\ne0$ (and $\psi$ used to calculate the overlap with the Jack state is in fact an excited state).
Symbols represent the points of highest overlap of $\psi$ with the Jack state (diamond labeled ``J''; this is simply the maximum of the displayed map) or Coulomb wave functions (open and full dots labeled ``0'' for LL$_0$ and ``1'' for LL$_1$ in GaAs, and square labeled ``2g'' for LL$_2$ in graphene i.e., G-LL$_2$; these maxima were determined from analogous maps of overlaps with those specific Coulomb states, like those in Fig.~\ref{fig3}).
More details are described in the main text.}
\label{fig2}
\end{figure}

\begin{figure}
\includegraphics[width=0.23\textwidth]{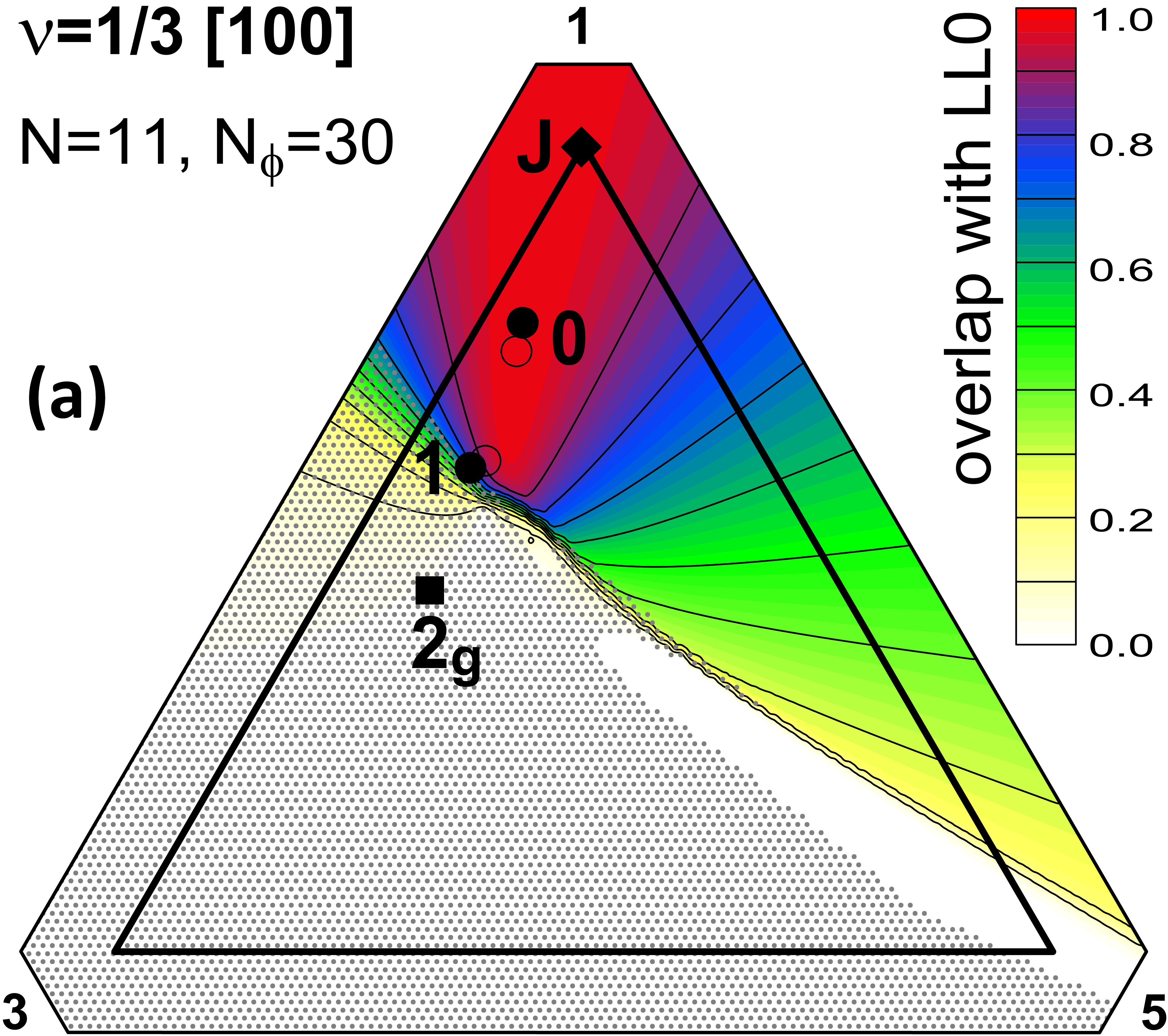}
\rule{0.01\textwidth}{0mm}
\includegraphics[width=0.23\textwidth]{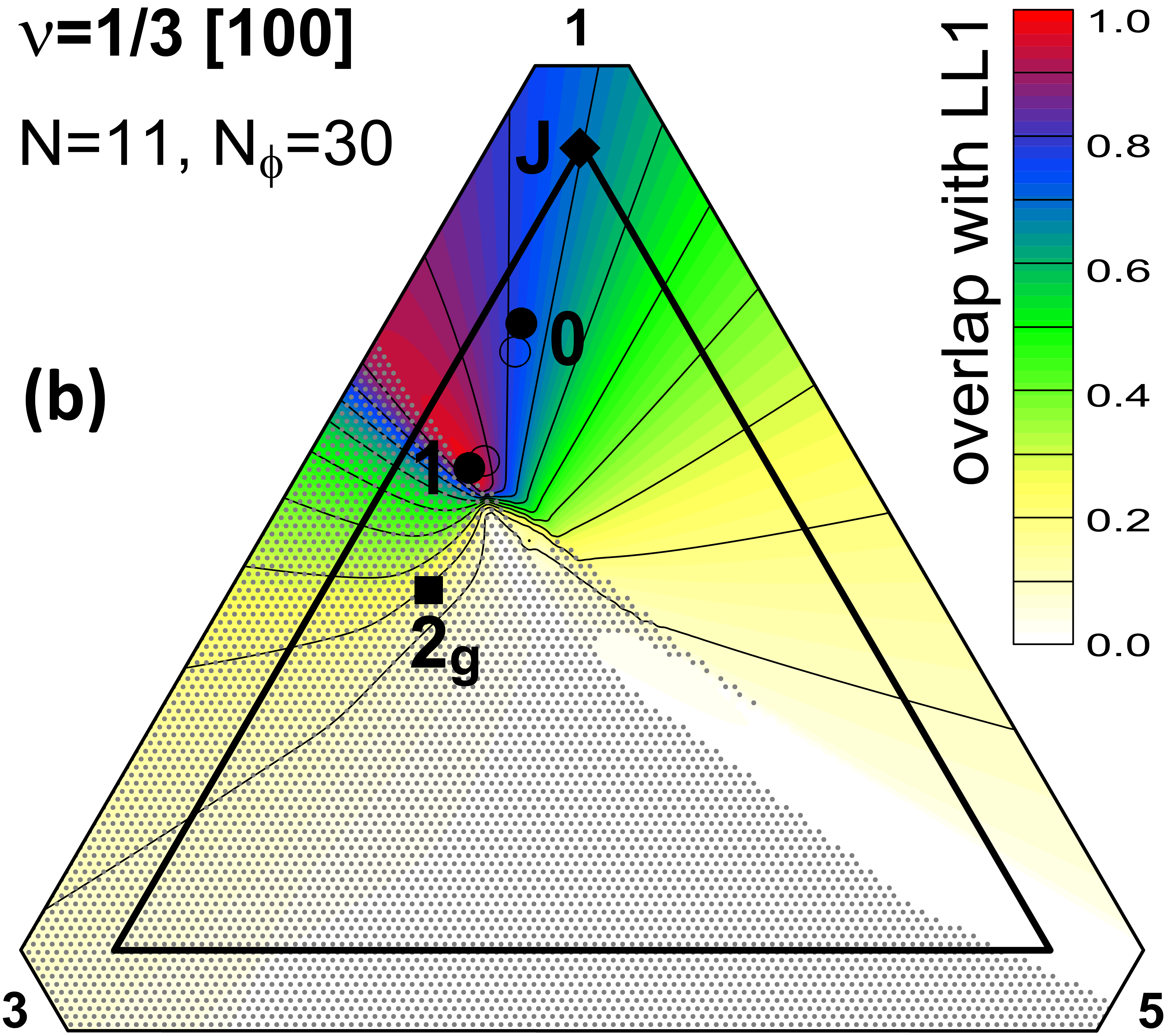}
\caption{
Maps similar to Fig.~\ref{fig2} and for the same system of $N=11$ electrons at flux $2Q=30$ but showing overlaps of $\psi(V_1,V_3,V_5)$ with the Coulomb ground states in a zero-width ($w=0$) GaAs layer in two different LLs: (a) LL$_0$, (b) LL$_1$.}
\label{fig3}
\end{figure}

In Fig.~\ref{fig2}, we have plotted a color map for the Jack state $[100]$, equivalent to the Laughlin $\nu=1/3$ state, and generated as a unique zero-energy ground state of the two-body pseudopotential with one non-vanishing (positive) coefficient, $V_1$, and all others vanishing.

This particular map corresponds to $N=11$ and $2Q=30$, which is the largest size we have for Jack state $[100]$; the maps for smaller sizes are similar so they have not been shown (for the same reason also for the other Jack states discussed in the following sections we will only show the maps for the largest available systems).
In the map, color contours indicate the overlap of the Jack state with the lowest-energy uniform ($L=0$) eigenstate of the model Hamiltonian $(V_1,V_3,V_5)$ called $\psi(V_1,V_3,V_5)$
The area of the map in which the ground state is non-uniform/degenerate (i.e., has $L>0$) has been marked by small grey dots (which also coincide with the computational grid used to calculate the map); in this dotted area the the overlapped $L=0$ model eigenstate lies above an unspecified lower state with $L>0$; only outside of this area (i.e., in the undotted part of the map) the overlapped $L=0$ model state is the absolute ground state.

The black diamond symbol labeled ``J'' indicates the point of maximum overlap, which in this case of course falls exactly at the $(1,0,0)$ corner of the map, where the generating Hamiltonian and the model are identical (hence, the answer to question Q1 is trivially positive for Jack state $[100]$).

The full and open black dots labeled ``0'' locate maximum overlaps of $\psi(V_1,V_3,V_5)$ with the Coulomb ground states in the lowest LL (LL$_0$) of massive fermions (e.g., in GaAs), for two extreme layer widths $w/l_B=0$ and 10, respectively (with the intermediate widths forming an unmarked continuous trace connecting the two dots).
Similarly, the two dots labeled ``1'' locate maximum overlaps with the Coulomb ground states in the second LL (LL$_1$) of massive fermions, for $w/l_B=0$ and 10.

For Dirac fermions (e.g., in graphene) we have only considered an ideal 2D layer with $w=0$.
Different LLs of graphene are denoted by G-LL$_n$.
However, as the Coulomb pseudopotentials in LL$_0$ and G-LL$0$ are identical, so the maximum overlap for G-LL$_0$ falls at the same point ``0g'' $\equiv$ ``0'' and has not been separately marked.
In G-LL$_1$ the Coulomb pseudopotential is slightly softer at short range than in LL$_0$, but still sufficiently strong to produce an essentially identical (upon mapping onto the lowest LL) $\nu=1/3$ ground state. 
So again, the maximum overlap for G-LL$_1$ falls at almost exactly the same point as in LL$_0$, ``1g'' $\equiv$ ``0'', and has not been separately marked.
Only for $n>1$ are the Coulomb ground states in graphene different and fall at different points on the map, for example the black square for $n=2$ has been explicitly labeled as ``2g''.
Also in all following figures the three Coulomb points for LL$_0$ ($w=0$), G-LL$_0$, and G-LL$_1$ coincide at the point collectively labeled ``0'', so the equivalent labels ``0g'' and ``1g'' will be ommitted.

\begin{table}
\centering
\begin{tabular}{|c|c|c||c|c|c||c|}
\hline
material & $n$ & $w/l_B$ & $V_1$ & $V_3$ & $V_5$ & overlap \\
\hline
\hline
\multirow{6}{*}{GaAs} &
\multirow{3}{*}{0}
             &  0 & 0.782 & 0.172 & 0.046 & 0.9999 \\ \cline{3-7}
         &   &  5 & 0.759 & 0.188 & 0.053 & 0.9998 \\ \cline{3-7}
         &   & 10 & 0.747 & 0.196 & 0.057 & 0.9997 \\ \cline{2-7}
         &
\multirow{3}{*}{1}
             &  0 & 0.602 & 0.317 & 0.081 & 0.9934 \\ \cline{3-7}
         &   &  5 & 0.609 & 0.302 & 0.089 & 0.9935 \\ \cline{3-7}
         &   & 10 & 0.611 & 0.297 & 0.092 & 0.9929 \\ \hline
\hline
\multirow{3}{*}{graphene} &
           0 &
\multirow{3}{*}{0}
                  & 0.782 & 0.172 & 0.046 & 0.9999 \\ \cline{2-2}\cline{4-7}
         & 1 &    & 0.777 & 0.178 & 0.045 & 0.9999 \\ \cline{2-2}\cline{4-7}
         & 2 &    & 0.450 & 0.437 & 0.113 & 0.9719 \\ \hline
\end{tabular}
\caption{
Locations and values of maximum overlaps between the indicated Coulomb ground states in GaAs and graphene and the lowest $L=0$ eigenstates $\psi(V_1,V_3,V_5)$ of the model pseudopotential, at filling factor $\nu=1/3$, for the system of $N=11$ electrons at flux $2Q=30$.}
\label{tab1}
\end{table}

While the dots and squares only show the points of maximum overlap, we have calculated full maps of the overlaps between each relevant Coulomb ground state and the model ground states $\psi(V_1,V_3,V_5)$.

For example, Figs.~\ref{fig3}(a) and \ref{fig3}(b) show the color contours for massive fermions in LL$_0$ and LL$_1$ (with the points of maximum overlap: ``J'', ``0'', ``1'', and ``2g'' of course the same as in Fig.~\ref{fig2}).
As already mentioned, the short-range model with the suitable choice of $V_1$, $V_3$, and $V_5$ is able to reproduce all considered Coulomb ground states with very high accuracy (see Tab.~\ref{tab1}), so the dots and squares in all maps can be considered as representing the exact Coulomb points (rather than as approximations limited by the $m\le5$ model).

It is remarkable (but of course not surprising) that Coulomb points ``0'', ``1'', and ``2g'' fall so far apart in the map, while the finite width moves them so relatively little (again, see Tab.~\ref{tab1}; their placement relative to ``monotonic'' and ``harmonic'' lines has also been indicated in Fig1.~\ref{fig1}).
However, when matching the Jack state with the Coulomb ground states via the maps $(V_1,V_3,V_5)$, it must be realized that it is always a whole area of high model/Jack or model/Coulomb match, extending around the indicated maximum point.
Since for the model/Coulomb match both the maximum point and the surrounding countour plot are very similar for any considered $\nu$ and $N$, we will not show them for other cases.

\subsubsection{Jack state $[10000]$ (Laughlin 1/5)}
\label{Jack10000}

\begin{figure}
\includegraphics[width=0.40\textwidth]{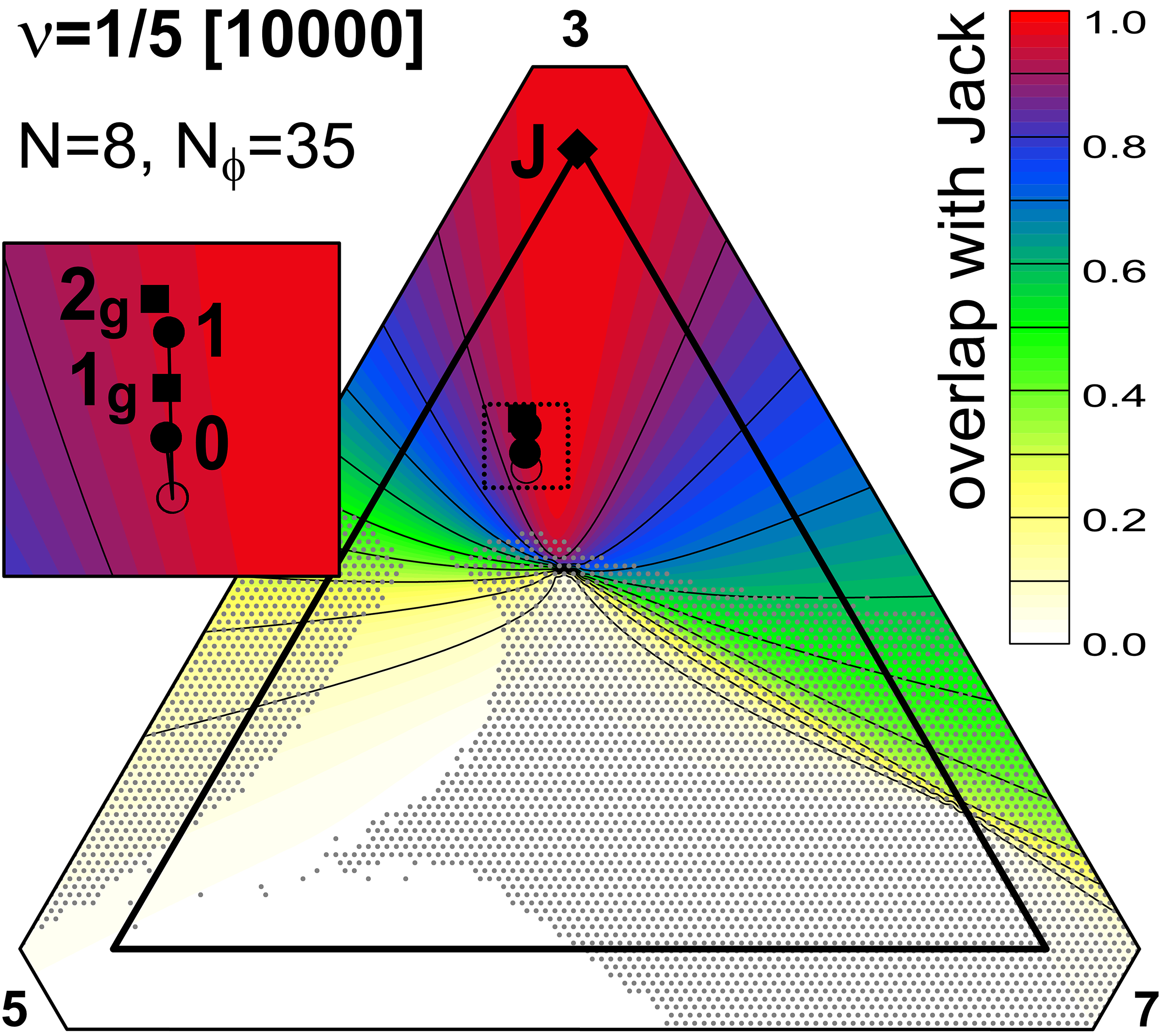}
\caption{
Map similar to Fig.~\ref{fig2} for $N=8$ electrons at flux $2Q=35$, showing overlaps of $\psi(V_3,V_5,V_7)$ with Jack state [10000] (i.e., the Laughlin $\nu=1/5$ state, generated by two-body repusion at $m=1$ and 3). 
Note that we used $V_1=\infty$ for this map with the three corners representing $m=3$, 5, and 7. 
The additional square panel shows the enlarged part indicated in the triangular map.}
\label{fig4}
\end{figure}

Jack state [10000], equivalent to the Laughlin $\nu=1/5$ state, is the unique zero-energy ground state of the two-body pseudopotential with positive both $V_1$ and $V_3$, and vanishing all other coefficients.
So compared to Jack state [100] from the previous section, it is still a two-body generating interaction, but with range extended to the next value of $m$.
Below filling factor $\nu=1/4$, all considered Coulomb ground states have negligible amplitude at pair angular momentum $m=1$, so we have calculated the map in coordinates $(V_3,V_5,V_7)$, corresponding to a modified short-range model with $V_1=\infty$, varied three coefficients at $m=3$, 5, and 7, and $V_m=0$ for $m>7$.
All Coulomb points for $\nu=1/5$ are essentially exact in this model, similarly as it was for $\nu=1/3$ and $(V_1,V_3,V_5)$ in Tab.~\ref{tab1}.

The overlap map for Jack state [10000] is shown in Fig.~\ref{fig4}, for $N=8$ and $2Q=35$.
The ``J'' point is exact at the top corner: $(V_3,V_5,V_7)=(1,0,0)$, and all Coulomb points lie close to one another, all in the red area of high overlap with the Jack state.
For LL$_0$ and LL$_1$ this confirms an earlier observation in, e.g., Fig.~2 of Ref.~\onlinecite{Wojs09a}.
The dotted rectangular part of the map containing all Coulomb points has been magnified to better show relative placement.

\subsubsection{Jack state $[1100]$ (Pfaffian 1/2)}

\begin{figure}
\includegraphics[width=0.40\textwidth]{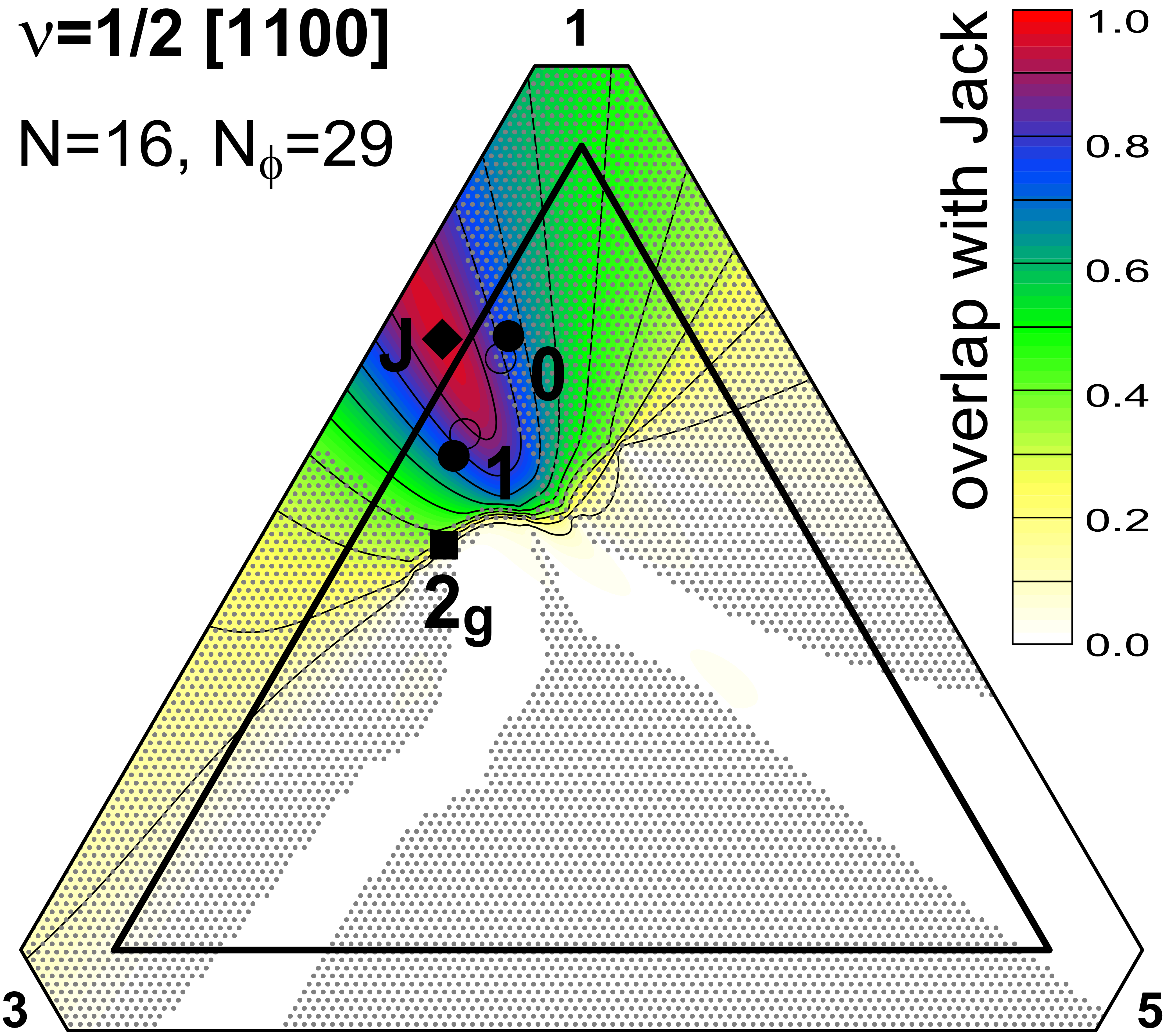}
\caption{
Map similar to Fig.~\ref{fig2} for $N=16$ electrons at flux $2Q=29$, showing overlaps of $\psi(V_1,V_3,V_5)$ with Jack state [1100] (i.e., the Pfaffian $\nu=1/2$ state, generated by three-body repusion at $m=3$).}
\label{fig5}
\end{figure}

Jack state [1100], equivalent to the Moore-Read ``Pfaffian'' $\nu=1/2$ state, is the unique zero-energy ground state of the three-body pseudopotential with a single non-zero (positive) coefficient at the relative triplet angular momentum $m=3$.

In general, the relative $K$-body angular momentum takes on values $m=m_{\rm min}$, $m_{\rm min}+2$, $m_{\rm min}+3$, \dots, where the minimum value is $m_{\rm min}=K(K-1)/2$.
For $K>2$, the $K$-body amplitudes and the corresponding $K$-body pseudopotentials $V^{(K)}_m$ are uniquely defined only up to a certain $m_{\rm max}$ (e.g., $m_{\rm max}=8$ for $K=3$), above which multiple states at the same $m$ exist, and $V^{(K)}_m$ becomes a matrix.
Nonetheless, while in this work we have not considered $K$-body pseudopotentials extending beyond $m_{\rm max}$, a model $K$-body interaction which is repulsive at one or more leading values of $m$ and vanishing for the higher ones can be defined regardless of the dimension of $V^{(K)}_m$.
Note that we have now added superscript $(K)$ to $V_m$, but with convention that $V^{(2)}\equiv V$, so that the notation used so far also holds.

The overlap map for Jack state [1100] is shown in Fig.~\ref{fig5}, for $N=16$ and $2Q=29$.
The ``J'' point is not exact, but almost so, with the overlap reaching 0.971 (see Tab.~\ref{tab2}).
Interestingly, it has one negative coordinate, but the red area of high Jack/model overlap reaches inside the positive triangle.
The positions and width dependencies of the Coulomb points ``0'' and ``1'' confirms the known fact that Jack state [1100] (Pfaffian, $p_x\pm ip_y$ superfluid of paired composite fermions) is a likely valid description of the half-filled LL$_1$, with the match improved by a finite width, while in LL$_0$ the Coulomb points fall into the dotted area of $L>0$ indicating compressibility (indeed, the half-filled LL$_0$ is a composite fermion Fermi sea).
Also the ``2g'' Coulomb point falls in the dotted (and low overlap) area, precluding emergence of Jack state [1100] in the half-filled G-LL$_2$.

\subsubsection{Jack state $[11000]$ (Gaffnian 2/5)}

\begin{figure}
\includegraphics[width=0.40\textwidth]{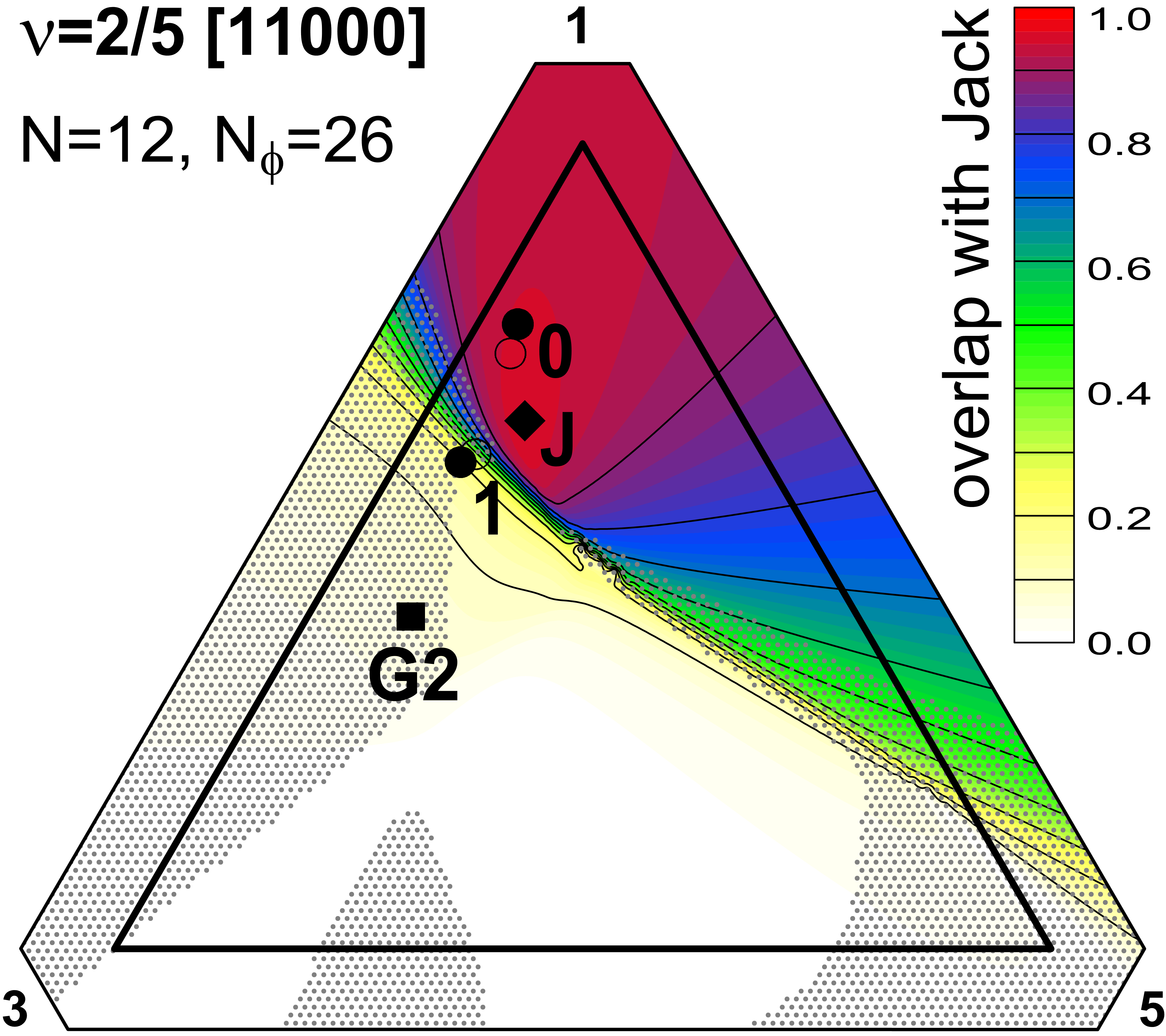}
\caption{
Map similar to Fig.~\ref{fig2} for $N=12$ electrons at flux $2Q=26$, showing overlaps of $\psi(V_1,V_3,V_5)$ with Jack state [11000] (i.e., the Gaffnian $\nu=2/5$ state, generated by three-body repusion at $m=3$ and 5).}
\label{fig6}
\end{figure}

Jack state [11000], equivalent to the ``Gaffnian'' $\nu=2/5$ state\cite{Simon07c}, is generated by the three-body pseudopotential with two positive coefficients at $m=3$ and 5, and all others vanishing.
Its overlap map is shown in Fig.~\ref{fig6}, for $N=12$ and $2Q=26$.

The ``J'' point lies now inside the triangle (see Tab.~\ref{tab2}), and the maximum overlap has the same value of 0.971 as for the Pfaffian.
The ``1'' and ``2g'' Coulomb points lie in the low overlap and $L>0$ areas, but the placement of the ``0'' point might suggest that Jack state [11000] (Gaffnian) is an accurate description of the $\nu=2/5$ state in the lowest LL.
However, this is known\cite{Toke09} to be an artifact of the finite size: in finite systems, Gaffnian and Jain $\nu=2/5$ states have high overlaps with each other and with the Coulomb ground state, but the two models are not equivalent and in fact they describe distinct topological orders in an infinite system, with the Jain state (of two filled composite fermion LLs) offering the proper description.

\subsubsection{Jack state $[110000]$ (Haffnian 1/3)}

\begin{figure}
\includegraphics[width=0.40\textwidth]{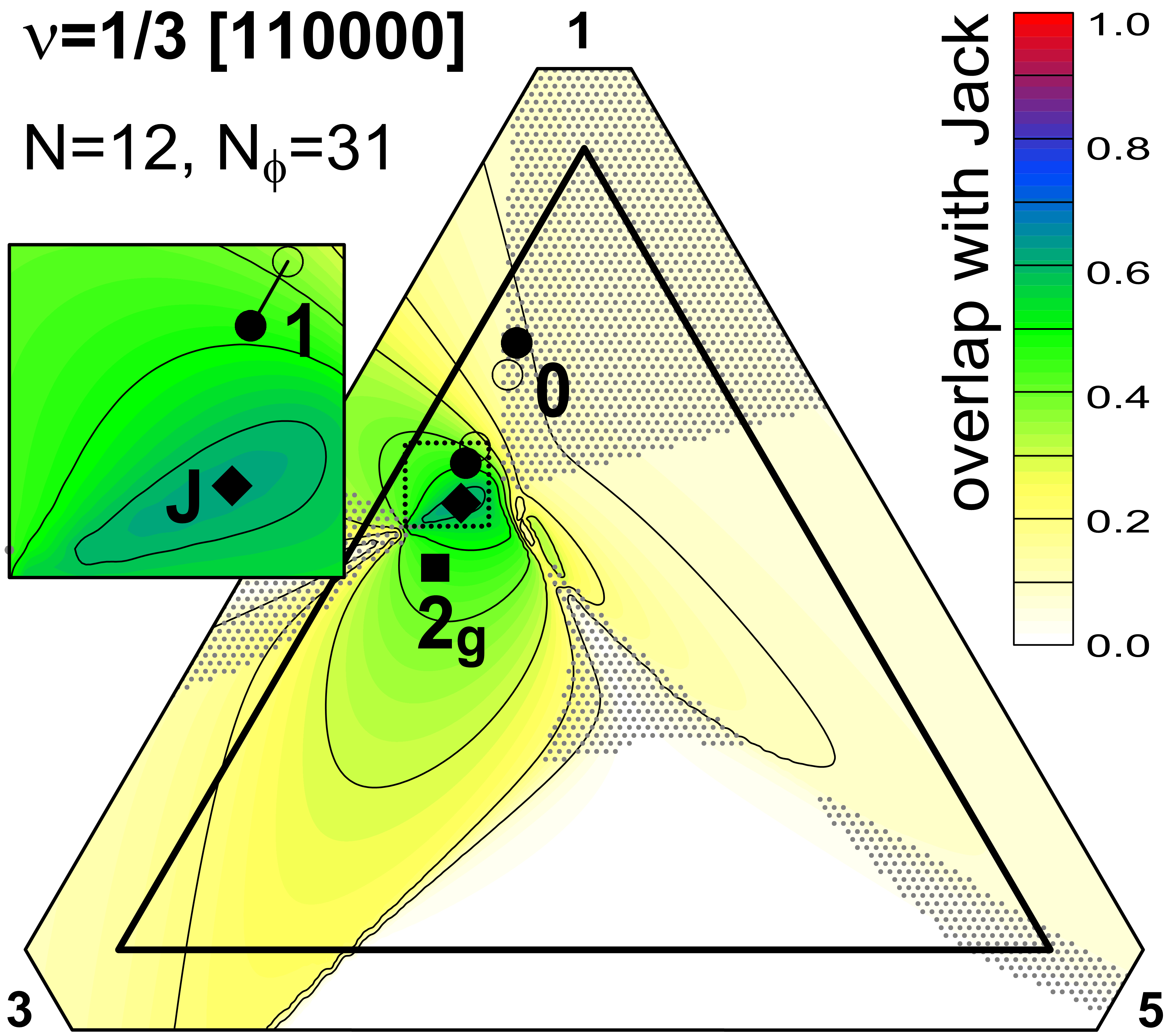}
\caption{
Map similar to Fig.~\ref{fig2} for $N=12$ electrons at flux $2Q=31$, showing overlaps of $\psi(V_1,V_3,V_5)$ with (improper) Jack state [110000] (i.e., the Haffnian $\nu=1/3$ state, generated by three-body repusion at $m=3$, 5 and 6).
The additional square panel shows the enlarged part indicated in the triangular map.}
\label{fig7}
\end{figure}

The ``Haffnian'' $\nu=1/3$ state\cite{Simon07b,Green01,Hermanns11} generated by the three-body pseudopotential with three positive coefficients at $m=3$, 5, and 6, and all others vanishing, corresponds to the fermionic Jack polynomial with root partition [110000] and $\alpha_{2,4}=-1$ which has a pole, and hence is not a proper Jack state.
Nonetheless, it can still be attributed root occupation and has been included in our analysis.
Its overlap map is shown in Fig.~\ref{fig7}, for $N=12$ and $2Q=31$.
In contrast to Pfaffian or Gaffnian, the maximum model/Haffnian overlap reaches a relatively low value of 0.63 at the ``J'' point $(V_1,V_3,V_5)=(0.56,0.35,0.09)$.
Remarkably, much higher overlaps are reached in smaller systems: 0.93 at point $(0.52,0.33,0.15)$ for $N=10$ and $2Q=25$, and 0.97 at point $(0.55,0.31,0.14)$ for $N=8$ and $2Q=19$, and 0.998 at point $(0.56,0.30,0.14)$ for $N=6$ and $2Q=13$.

\subsubsection{Jack state $[11100]$ (Parafermion 3/5)}

\begin{figure}
\includegraphics[width=0.40\textwidth]{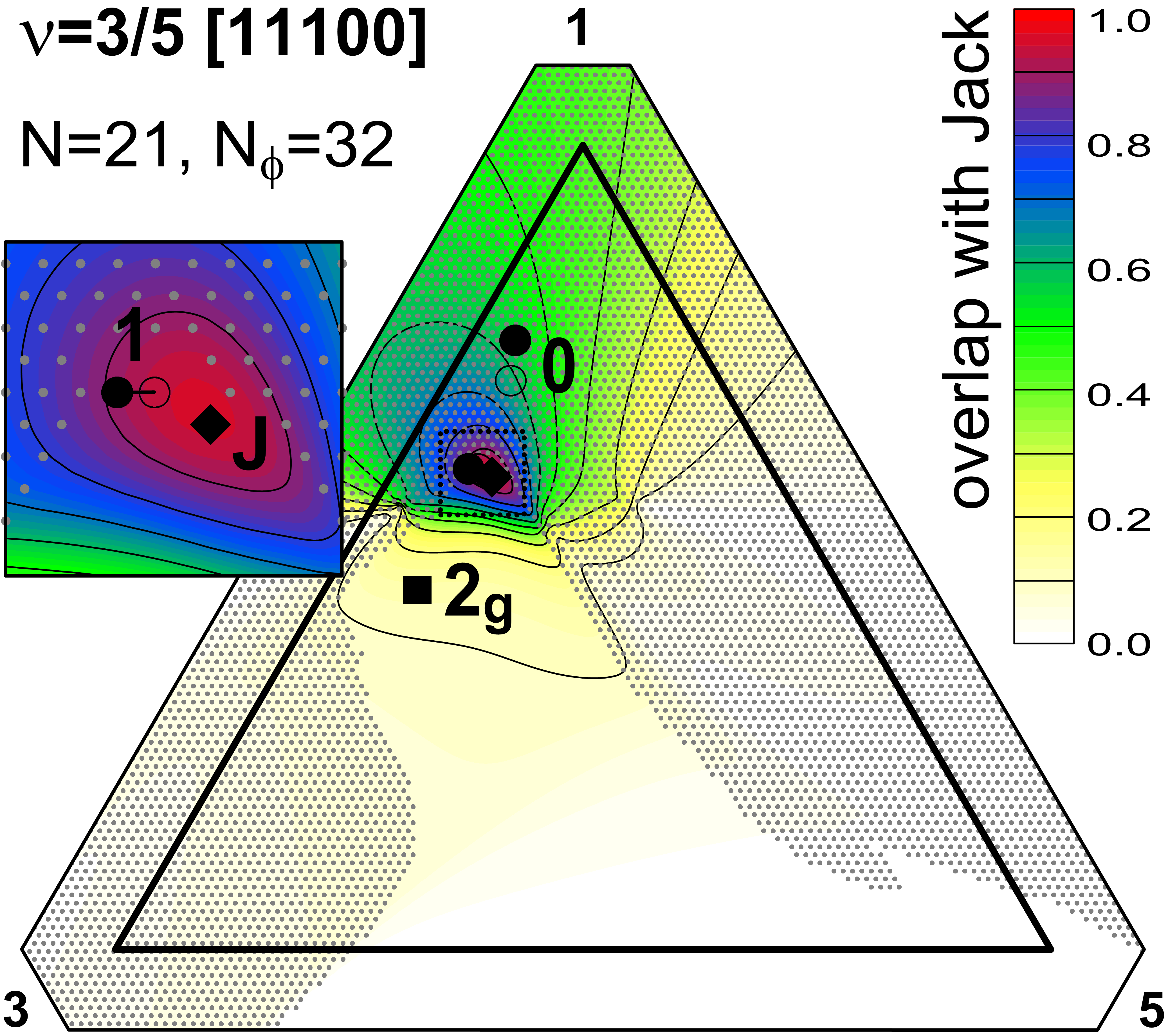}
\caption{
Map similar to Fig.~\ref{fig2} for $N=21$ electrons at flux $2Q=32$, showing overlaps of $\psi(V_1,V_3,V_5)$ with Jack state [11100] (i.e., the parafermion $\nu=3/5$ state, generated by four-body repusion at $m=6$).
The additional square panel shows the enlarged part indicated in the triangular map.}
\label{fig8}
\end{figure}

Jack state [11100], equivalent to the Read-Rezayi ``parafermion'' $\nu=3/5$ state\cite{Read99}, is generated by the four-body pseudopotential with one positive coefficients at the smallest relative four-body angular momentum $m=6$, and all others vanishing.
Its overlap map is shown in Fig.~\ref{fig8}, for $N=21$ and $2Q=32$.

The ``J'' point lies now inside the triangle, and the maximum overlap has a high value of 0.968 (see Tab.~\ref{tab2}).
It may be worth stressing that for this Jack state (like for all Jack states for which it is not clearly stated otherwise) both the position and overlap of the ``J'' point are very similar in smaller systems (we have also checked $N=18$ at $2Q=27$ and $N=15$ at $2Q=22$).
Remarkably, the ``J'' point is surrounded by a rather small (compared to other Jack states) undotted area of $L=0$, which however securely includes both ``J'' and ``1'' points (as clearly seen in the inset showing the relevant part of the map in magnification).
It is also evident that increasing layer width $w$ of the Coulomb system improves the match of the Jack and Coulomb ($n=1$) states.
This observation is consistent with earlier analysis\cite{Read99,Simon07a,Wojs09a,Gurarie00} of energies pointing to Jack state [11100] as the most likely description of the $\nu=13/5$ (and, by particle-hole conjugation, $\nu=12/5$) FQH state in GaAs.
A new conclusion is that Jack state [11100] is unlikely to emerge in graphene (in any LL).

\subsubsection{Jack state $[111100]$ (Parafermion 2/3)}

\begin{figure}
\includegraphics[width=0.40\textwidth]{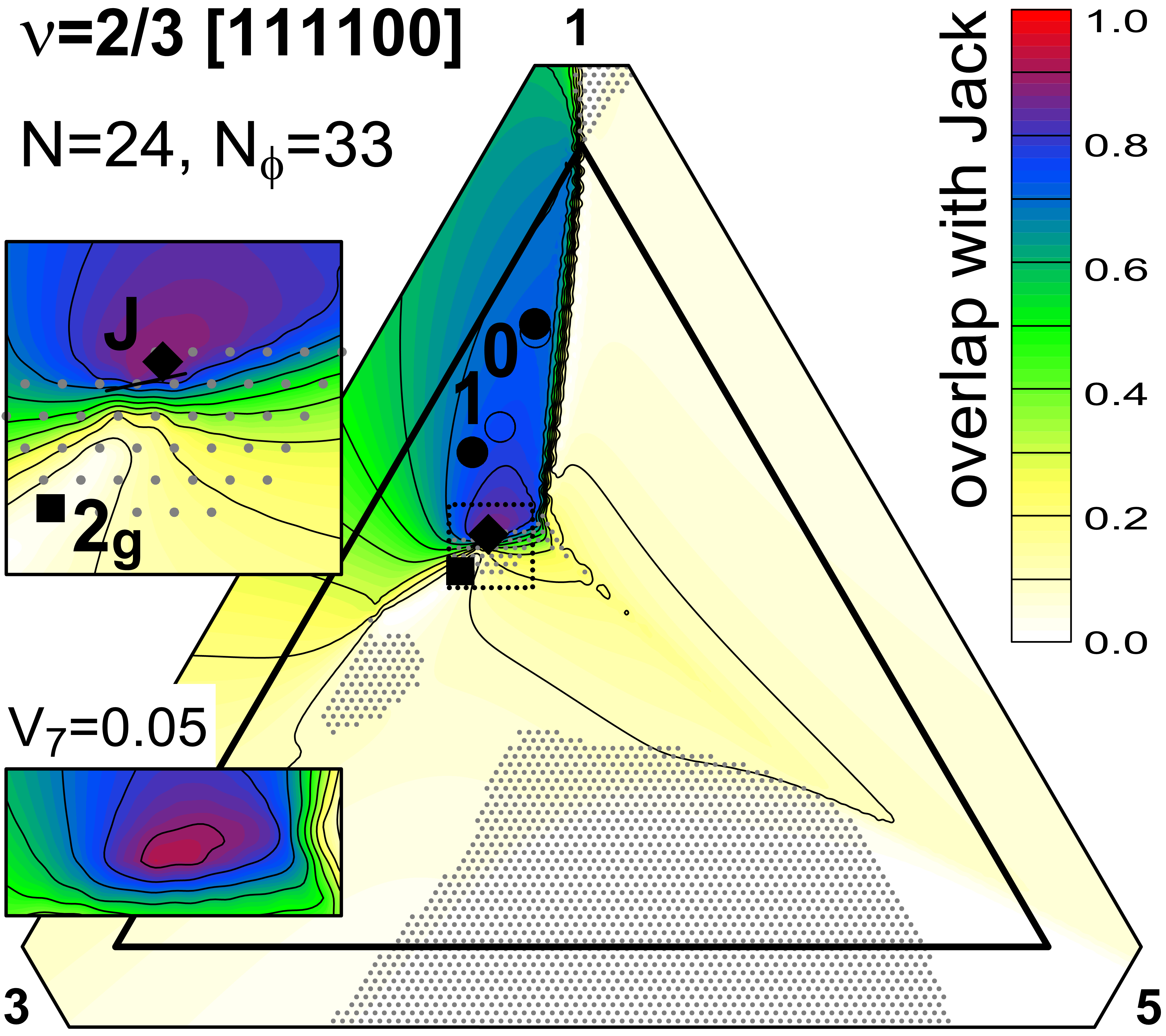}
\caption{
Map similar to Fig.~\ref{fig2} for $N=24$ electrons at flux $2Q=33$, showing overlaps of $\psi(V_1,V_3,V_5)$ with Jack state [111100] (i.e., the parafermion $\nu=2/3$ state, generated by five-body repusion at $m=10$). 
The upper additional square panel shows the enlarged part indicated in the triangular map.
The lower additional rectangular panel shows part of the $(V_1,V_3,V_5)$ map for $V_7=0.05$ (see explanation in main text).}
\label{fig9}
\end{figure}

Jack state [111100], equivalent to the Read-Rezayi ``parafermion'' $\nu=2/3$ state\cite{Read99}, is generated by the five-body pseudopotential with one positive coefficients at the smallest relative four-body angular momentum $m=10$, and all others vanishing.
Its overlap map is shown in Fig.~\ref{fig9}, for $N=24$ and $2Q=33$.

When the search for maximum overlap is limited to the $(V_1,V_3,V_5)$ plane, its value at the optimum ``J'' point is the unimpressive 0.896.
Moreover, the ``J'' point falls into the dotted area, meaning that the pair Hamiltonian best reproducing this Jack state has a lower state at $L>0$.

However, we are guided by observation that accurate reproduction of Jack states [100], [1100], and [11100] by a pair Hamiltonian requires (suitable) repulsion at $m=1$ (corner of the triangle), $m=1$ and 3 (edge of the triangle), and $m=1$, 3 and 5 (inside of the triangle).
Thus, we can anticipate that adding variable $V_7$ to the search space should lift the model/Jack overlap close to unity.
With normalization $V_1+V_3+V_5+V_7=1$, this corresponds to a search for an optimum match inside a tetrahedron (pyramid) with a base corresponding to $V_7=0$.
Indeed, a considerably higher overlap of 0.945 is reached inside the pyramid, at $(V_1+V_3+V_5+V_7)=(0.49,0.32,0.14,0.05)$.
The relevant part of the $V_7=0.05$ section of the 3D overlap map is shown as inset in Fig.~\ref{fig9};
its scale is the same as of the main ($V_7=0$) map and the maxima lie almost exactly one above the other.
Not only has the overlap increased when going from $V_7=0$ to 0.05, but also the $L=0$ (undotted) area has greatly expanded, including the whole shown area of the map.
We will return to these facts in the next Sec.~\ref{general}.

The location of the Coulomb points on the map suggests that Jack state [111100] is possible to emerge in LL$_1$.
This may seem as an attractive hypothesis to explain the $\nu=7/3$ FQHE, and its weakness compared to $\nu=1/3$ in the lowest LL.
However, the $\nu=7/3$ state has already recently been explained\cite{Balram13} as Laughlin state with strong composite fermion excitonic effects, so the relevance of Jack state [111100] is doubtful (although further studies aimed specifically at this problem might be interesting).
On the other hand, our map suggests that Jack state [111100] is unlikely to form in LL$_0$ in GaAs or in any LL in graphene.

\subsubsection{General result for $K$-body contact repulsion}
\label{general} 

The above state-by-state analysis suggests a general relation between the order $K$ of the contact interaction (defined by the $K$-body pseudopotential with only a single non-negative and positive coefficient at $m=m_{\rm min}\equiv K(K-1)/2$) and the range of model pair interaction able to accurately reproduce the same (Jack) ground state: The pair Hamiltonian must have suitable positive coefficients at $m<2K-1$.
The sequence of pair pseudopotentials most accurately generating Jack states [100], [1100], [11100], and [111100] based on our overlap maps has been listed (along with the overlaps) in Tab.~\ref{tab2}.
These pseudopotentials have been optimized only at $m<2K-1$, with higher coefficients set to zero, i.e., best fits to [100], [1100], [11100], and [111100] are searched at the corner, side, base, and in the whole tetrahedron of the $(V_1,V_3,V_5,V_7)$ model.
For each Jack state we used the map for the largest system available.

\begin{table}
\centering
\begin{tabular}{|r|c||c|c|c|c|c||c|}
\hline
Jack & $K$ & $V_1$ & $V_3$ & $V_5$ & $V_7$ & $V_9$ & overlap \\
\hline
\hline
   [100] & 2 & 1.00 & 0    & 0    & 0    & 0    & 1     \\ \hline
  [1100] & 3 & 0.73 & 0.27 & 0    & 0    & 0    & 0.968 \\ \hline
 [11100] & 4 & 0.59 & 0.30 & 0.11 & 0    & 0    & 0.968 \\ \hline
[111100] & 5 & 0.49 & 0.32 & 0.14 & 0.05 & \phantom{0}0\phantom{0} & 0.945 \\ \hline
\end{tabular}
\caption{
Pair pseudopotentials $V_m$ whose ground states have maximum overlap with the indicated series of Jack states, generated as unique zero-energy ground states of $K$-body contact repulsion (corresponding to the $K$-body pseudopotential with a positive single leading coefficient and all others vanishing).
For each $K$, only $V_m$ at $m<2K-1$ were optimized and higher coefficients were set to zero.
The systems sizes used in the calculation are:
[100] -- any size (result is exact);
[1100] -- $N=16$ and $2Q=29$;
[11100] -- $N=21$ and $2Q=32$;
[111100] -- $N=24$ and $2Q=33$.}
\label{tab2}
\end{table}

Inspection of Tab.~\ref{tab2} reveals that the ratios of consecutive pseudopotential coefficients are (to an excellent approximation) $V_1$:$V_3$:$V_5$:$V_7=1$:0:0:0, 3:1:0:0, 6:3:1:0, and 10:6:3:1 for $K=2$, 3, 4, and 5.
In Tab.~\ref{tab3} we list overlaps calculated for systems of different sizes $N$ for pair pseudopotentials defined by this simple regularity, i.e., given by (apart from the irrelevant normalization):
\begin{equation}
V^{(K)}_m\sim(2K-1-m)(2K+1-m).
\label{eq1}
\end{equation}
All overlaps in Tab.~\ref{tab3} are nearly as high as in Tab.~\ref{tab2}, confirming validitity of the regularity and suggesting that it may also be valid for higher $K$'s.

It is noteworthy that for $K=3$ the proposed formula agrees with results of the recent paper discussing mean field approximation of three-body interactions\cite{Sreejith17}.

\begin{table}
\centering
\begin{tabular}{|r|c||c|c|c|c|c||c|}
\hline
Jack & $K$ & $V_1$ & $V_3$ & $V_5$ & $V_7$ & $V_9$ & overlap$_{(N)}$ \\
\hline
\hline
   [100] & 2 &  1 & 0 & 0 & 0 & 0 & 1$_{\rm (any)}$ \\ \hline
  [1100] & 3 &  3 & 1 & 0 & 0 & 0 & 0.949$_{(16)}$ 0.950$_{(14)}$ 0.945$_{(12)}$ \\ \hline
 [11100] & 4 &  6 & 3 & 1 & 0 & 0 & 0.967$_{(21)}$ 0.871$_{(18)}$ 0.971$_{(15)}$ \\ \hline
[111100] & 5 & 10 & 6 & 3 & 1 & 0 & 0.909$_{(24)}$ 0.964$_{(20)}$ 0.954$_{(16)}$ \\ \hline
\end{tabular}
\caption{
Table similar to Tab.~\ref{tab2} but for pair pseudopotentials defined by Eq.(\ref{eq1}) and overlaps given for systems of different size $N$ (indicated as subscript at each overlap).}
\label{tab3}
\end{table}

The above Eq.~\ref{eq1} and Tab.~\ref{tab3} express the main result of this work:
The ground state of a contact {\em many-body} ($K$-body) repulsion is accurately reproduced by a {\em two-body} pseudopotential with coefficients taken from the simple sequence: 1, 3, 6, 10, \dots.

\begin{figure}
\includegraphics[width=0.40\textwidth]{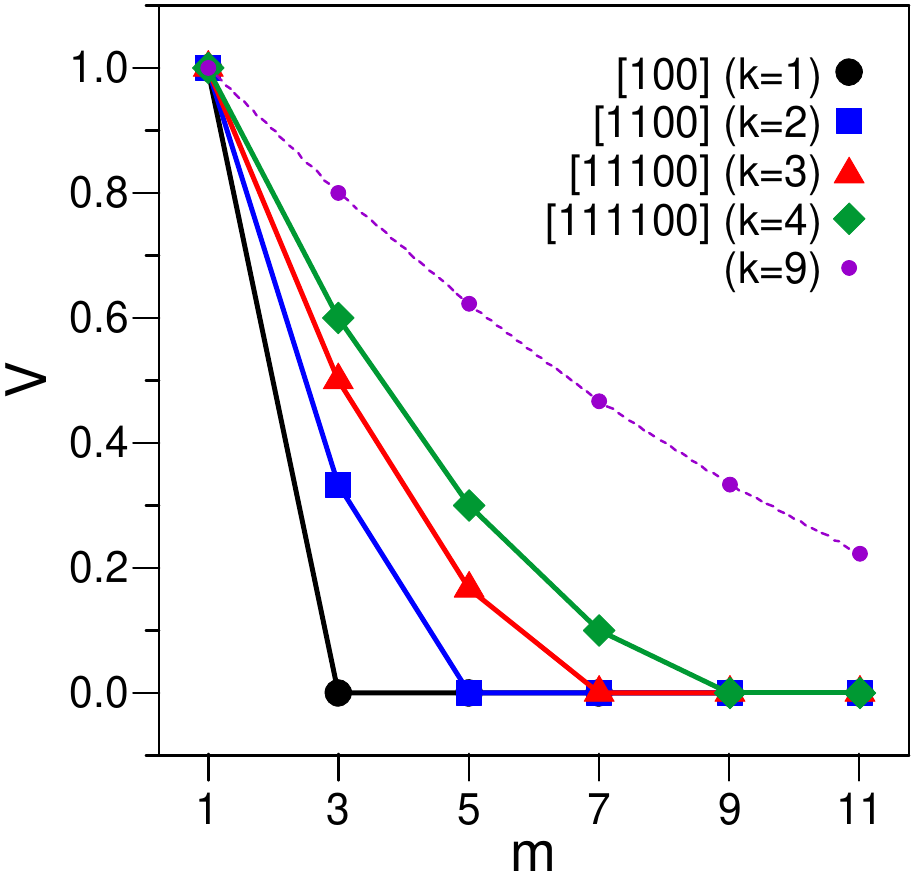}
\caption{
Pseudopotentials from Tab.~\ref{tab3}, normalized to $V_1\equiv1$, with an additional one for a rather high $K=9$ given by Eq.~(\ref{eq1}).}
\label{fig10}
\end{figure}

Several of these pseudopotentials have been plotted in Fig.~\ref{fig10}, normalized so that $V_1\equiv1$.
While the most the interesting dynamics (emergence of Jack ground states) is induced by these pseudopotentials at rather high filling factors $\nu=1-2/(K+1)$, it should also be noted that they are all superharmonic at each $m$ where they are positive, so they support formation of composite fermions and a series of Laughlin states at $\nu\ge(2K-1)^{-1}$.
However, their superharmonicity waekens with increasing $K$ (as is clearly seen for a large $K=9$), and for $m\ll K$ the pseudopotential (corresponding to an infinite-body contact repulsion) becomes linear in $m$:
\begin{equation}
V^{(K)}_{m\ll K}\sim1-{m-1\over K}
\end{equation}
i.e., harmonic, and as such it does not induce any correlations whatsoever.\cite{Wojs98,Wojs99,Wojs00,Wojs09b}

We have also noticed that the number of pair pseudopotential coefficients needed to accurately generate the same ground state as the $K$-body pseudopotential with $k\ge1$ positive coefficients also grows with increasing $k$.
A trivial example is the Laughlin $\nu=(2k+1)^{-1}$ series of Jack states with $K=2$, but our maps show the same effect for the Pfaffian-Gaffnian-Haffnian sequence with $K=3$.

\subsubsection{Jack states in Coulomb systems}

Summarizing our results regarding the possible emergence of Jack ground states in realistic systems of electrons interacting by Coulomb forces in an arbitrary ($n=0$, 1, \dots) LL in GaAs or graphene, we can state the following.
As is well-known, Jack states [100] and [10000] (i.e., Laughlin states at $\nu=1/3$ and $1/5$) are robust FQH states in LL$_0$ (in GaAs) and in G-LL$_0$ and G-LL$_1$ (in graphene).
Moreover, Jack state [10000] should also form in LL$_1$ and G-LL$_2$.
As is also well-known, Jack state [1100] (i.e., Pfaffian at $\nu=1/2$) should occur in LL$_1$ (and nowhere else).
Judging from our maps alone, Jack state [11000] (Gaffnian at $\nu=2/5$) might look like a good candidate in LL$_0$, but it is known\cite{Toke09} that it has higher energy than Jain's state of composite fermions filling two effective LLs.
Jack state [110000] (Haffnian at $\nu=1/3$) might seem like a possible candidate for the LL$_1$ (to explain FQHE in GaAs at $\nu=7/3$ or $8/3$) but earlier studies\cite{Wojs09a,Wojs01} have not confirmed a complete series of gapped $L=0$ Coulomb ground states at the corresponding flux $2Q=3N-5$.
Furthermore, Haffnian is not a proper Jack and it has been argued to be compressible\cite{Simon07b,Green01,Hermanns11}.
Jack state [11100] (parafermion at $\nu=3/5$) is likely to occur in LL$_1$ and thus to underlie FQHE at $\nu=12/5$ and $13/5$ in GaAs.
Finally, Jack state [111100] (parafermion at $\nu=2/3$) could occur in LL$_1$ and explain FQHE at $\nu=7/3$ and $8/3$ in GaAs, but certainly far more thorough studies would be needed to rival the current picture\cite{Balram13} of adiabatic connection to the Laughlin state at these fillings.

\begin{table}
\centering
\begin{tabular}{|r|c|c|c||c|c|c|c|c|}
\hline
Jack & $N$ & $2Q$ & dim & LL$_0$ & LL$_1$ & LL$_1^{\rm wide}$ & G-LL$_1$ & G-LL$_2$ \\
\hline
\hline

        [100] & 11 & 30 & $1\!\cdot\!10^6$ & 0.9922 & 0.7030 & 0.8199 & 0.9901 & 0.0093 \\ \cline{2-9}
$\nu\!=\!1/3$ & 12 & 33 & $8\!\cdot\!10^6$ & 0.9909 & 0.5030 & 0.7141 & 0.9885 & 0.0003 \\ \cline{2-9}
              & 13 & 36 & $4\!\cdot\!10^7$ & 0.9898 & 0.5445 & 0.7332 & 0.9871 & 0.0013 \\ \cline{2-9}
              & 14 & 39 & $3\!\cdot\!10^8$ & 0.9887 & 0.5771 & 0.7411 & 0.9858 & 0.0013 \\
\hline
\hline
      [10000] &  7 & 30 & $5\!\cdot\!10^4$ & 0.9768 & 0.9818 & 0.9776 & 0.9792 & 0.9800 \\ \cline{2-9}
$\nu\!=\!1/5$ &  8 & 35 & $4\!\cdot\!10^5$ & 0.9589 & 0.9678 & 0.9603 & 0.9631 & 0.9641 \\ \cline{2-9}
              &  9 & 40 & $4\!\cdot\!10^6$ & 0.9334 & 0.9453 & 0.9345 & 0.9388 & 0.9374 \\ \cline{2-9}
              & 10 & 45 & $4\!\cdot\!10^7$ & 0.9228 & 0.9386 & 0.9250 & 0.9302 & 0.9320 \\
\hline
\hline
       [1100] & 14 & 25 & $2\!\cdot\!10^5$ & 0.7223 & 0.6935 & 0.8155 & 0.7298 & 0.2584 \\ \cline{2-9}
$\nu\!=\!1/2$ & 16 & 29 & $2\!\cdot\!10^6$ & 0.7459 & 0.7795 & 0.8443 & 0.7517 & 0.0895 \\ \cline{2-9}
              & 18 & 33 & $3\!\cdot\!10^7$ & 0.6355 & 0.6766 & 0.7633 & 0.6410 & 0.1322 \\ \cline{2-9}
              & 20 & 37 & $4\!\cdot\!10^8$ & 0.3703 & 0.6736 & 0.7829 & 0.3756 & 0.1687 \\
\hline
\hline
      [11000] & 10 & 21 & $2\!\cdot\!10^3$ & 0.9715 & 0.2748 & 0.3326 & 0.9713 & 0.0369 \\ \cline{2-9}
$\nu\!=\!2/5$ & 12 & 26 & $3\!\cdot\!10^4$ & 0.9646 & 0.2119 & 0.2900 & 0.9642 & 0.0726 \\ \cline{2-9}
              & 14 & 31 & $7\!\cdot\!10^5$ & 0.9582 & 0.1600 & 0.2777 & 0.9574 & 0.0067 \\ \cline{2-9}
              & 16 & 36 & $1\!\cdot\!10^7$ & 0.9526 & 0.1096 & 0.2691 & 0.9516 & 0.0091 \\
\hline
\hline
     [110000] &  8 & 19 & $4\!\cdot\!10^3$ & 0.3131 & 0.6709 & 0.6194 & 0.3192 & 0.7220 \\ \cline{2-9}
$\nu\!=\!1/3$ & 10 & 25 & $1\!\cdot\!10^5$ & 0.1521 & 0.7205 & 0.7297 & 0.1515 & 0.6952 \\ \cline{2-9}
              & 12 & 31 & $3\!\cdot\!10^6$ & 0.1096 & 0.5182 & 0.4603 & 0.1107 & 0.4613 \\ \cline{2-9}
              & 14 & 37 & $1\!\cdot\!10^8$ & 0.0619 & 0.1074 & 0.0500 & 0.0623 & 0.5866 \\
\hline
\hline
      [11100] & 15 & 22 & $1\!\cdot\!10^4$ & 0.8315 & 0.9836 & 0.9801 & 0.8338 & 0.2060 \\ \cline{2-9}
$\nu\!=\!3/5$ & 18 & 27 & $2\!\cdot\!10^5$ & 0.5399 & 0.9369 & 0.8995 & 0.5458 & 0.3584 \\ \cline{2-9}
              & 21 & 32 & $5\!\cdot\!10^6$ & 0.5689 & 0.8990 & 0.9316 & 0.5714 & 0.1332 \\ \cline{2-9}
              & 24 & 37 & $1\!\cdot\!10^8$ & 0.3442 & 0.8100 & 0.8792 & 0.3468 & 0.1408 \\
\hline
\hline
     [111100] & 20 & 27 & $6\!\cdot\!10^4$ & 0.6186 & 0.8675 & 0.8563 & 0.6161 & 0.5082 \\ \cline{2-9}
$\nu\!=\!2/3$ & 24 & 33 & $2\!\cdot\!10^6$ & 0.7349 & 0.7697 & 0.7832 & 0.7358 & 0.1139 \\
\hline
\end{tabular}
\caption{
Overlaps of the indicated of Jack states with different Coulomb ground states in the zero angular momentum channel ($L=0$). Consecutive columns are: root occupation $[\dots]$ and filling factor $\nu$, electron number $N$, magnetic flux on the sphere $2Q$, dimension of the relevant $N$-body subspace with zero total angular momentum projection ($L_z=0$), and the overlaps with Coulomb states in the $n=0$ and 1 LLs in GaAs (LL$_n$) and in the $n=1$ and 2 LLs in graphene (G-LL$_n$). Layer width for each Coulomb system is zero, except for LL$_1^{\rm wide}$ corresponding to $w/l_B=3$.}
\label{tab4}
\end{table}

\begin{figure}
\includegraphics[width=0.48\textwidth]{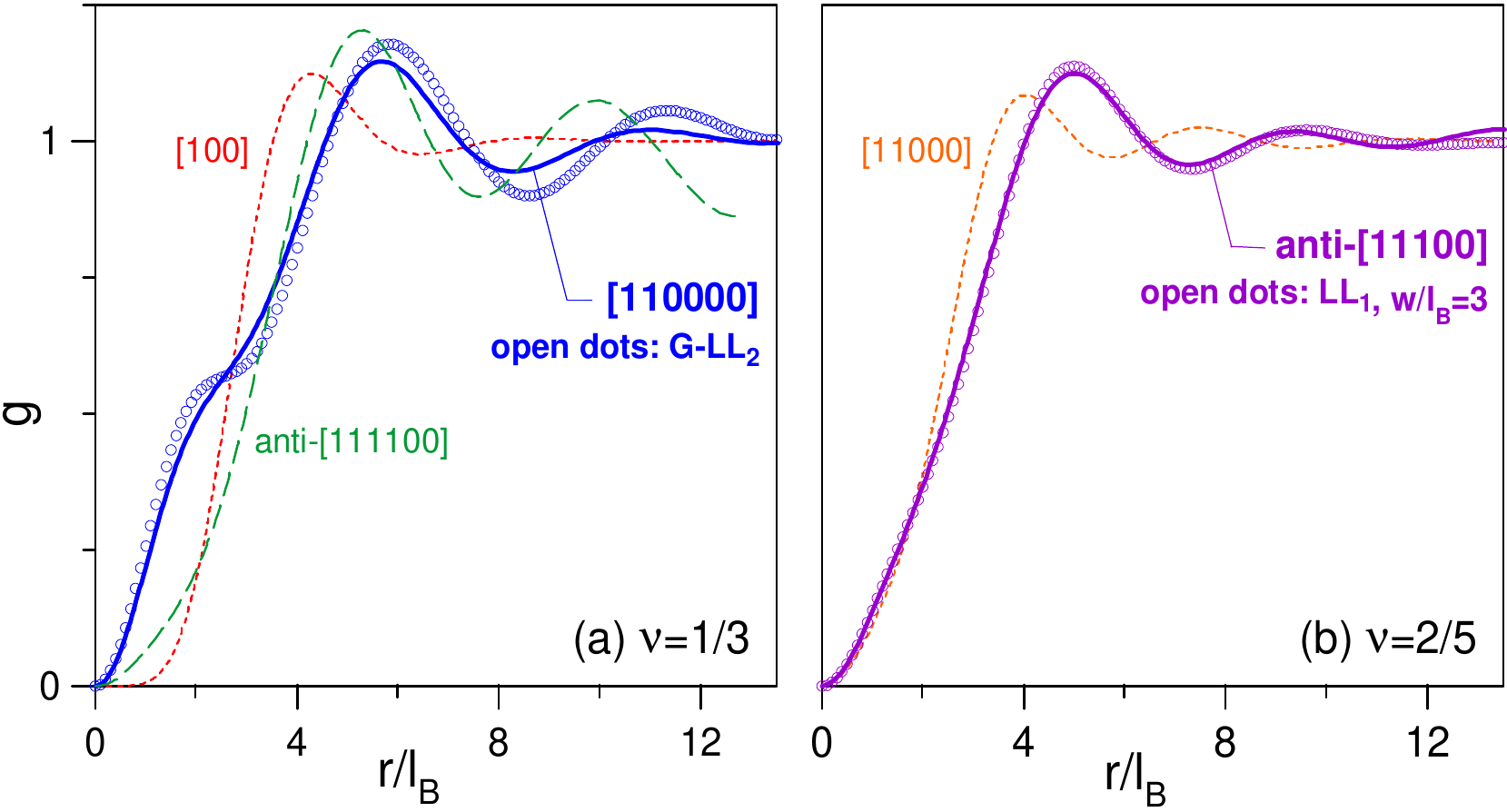}
\caption{
Pair correlation functions $g(r)$, with distance $r$ expressed in the units of magnetic length $l_B$, of select Jack and Coulomb ground states (or their particle-hole conjugates).
(a) $\nu=1/3$: dotted red curve -- Jack state [100] (Laughlin state) calculated for $N=14$ and $2Q=39$, blue solid curve and blue open dots -- (improper) Jack state [110000] (Haffnian) and Coulomb ground state in G-LL$_2$ for $N=14$ and $2Q=37$, green dashed curve -- conjugate of Jack state [111100] (anti-parafermion) for $N=10$ and $2Q=33$. 
(b) $\nu=2/5$: orange dotted curve -- Jack state [11000] (Gaffnian) for $N=16$ and $2Q=36$, purple solid curve and purple open dots -- conjugate of Jack state [11100] (anti-parafermion) and Coulomb ground state in LL$_1$ (for a fairly wide layer of $w/l_B=3$) for $N=14$ and $2Q=37$.}\label{fig11}
\end{figure} 

The list of overlaps in Tab.~\ref{tab4} is complemented with Fig.~\ref{fig11} showing comparison of pair correlation functions $g(r)$ of different Jack and Coulomb ground states (or their particle-hole conjugates). 
In the left panel (a) for $\nu=1/3$, it is well-known that Jack state [100] (Laughlin state) has almost the same correlations as the Coulomb ground state in LL$_0$ (not shown). 
But it is quite remarkable how accurately the (improper) Jack state [110000] (Haffnian) matches the Coulomb ground state in G-LL$_2$ (providing far stronger support for their connection than merely moderate overlaps of Tab.~\ref{tab4}). 
On the other hand, the conjugate of Jack state [111100] (anti-parafermion) shows strong long-range oscillations and is rather different from any considered Coulomb ground state. 
In the right panel (b) for $\nu=2/5$, it is well-known that Jack state [11000] (Gaffnian) has almost the same correlations as the Coulomb ground state in LL$_0$ (not shown) and that it is nonetheless topologically distinct from the composite fermion state which is known to offer a correct description for this Coulomb system. 
But it is remarkable how accurately the conjugate of Jack state [11100] (Read-Rezayi parafermion state) matches the Coulomb ground state in LL$_1$, especially in a sufficiently wide layer (in the figure we used $w/l_B=3$); in this case their apparent connection is also consistent with other evidence (overlaps and energies).

\section{Conclusions}
\label{sec4}

We examined a series of FQHE wave functions based on fermionic Jack polynomials which are also the ground states of particular short-range multi-particle repulsion. 
Our analysis revolved around the examined overlaps of trial wave functions and ground states of suitable two-body Hamiltonians. 
Our results reveal that Coulomb ground states (for both massive/Sch\"odinger electrons in GaAs and massless/Dirac electrons in graphene, and including their varation with the LL index and layer width) are represented with excellent accuracy by pair model pseudopotentials with only a few suitable leading coefficients. 
Jack states (or, in general, the ground states of short-range $K$-body repulsive interactions) are also reproduced with high accuracy by the short-range pair model. 
In particular, we found a simple formula (\ref{eq1}) for a two-body pseudopotential with $K-1$ leading coefficients which accurately reproduces Jack states [11\dots100] generated by the contact $K$-body repulsion. 
Options for finding Jack states in realistic Coulomb systems in GaAs or monolayer graphene are probably limited to the obvious Laughlin states and the commonly accepted Pfaffian and parafermion states.

\section{ACKNOWLEDGMENTS}

The authors are grateful to Jainendra K. Jain, Yinghai Wu, Ganesh Sreejith, Pawe{\l} Potasz, and Steven H. Simon for useful discussions. 
This work was supported by the Polish NCN Grant No. 2014/14/A/ST3/00654. 
The computations were largely done at Wroc{\l}aw Centre for Networking and Supercomputing and Academic Computer Centre CYFRONET, both parts of PL-Grid Infrastructure.

\end{document}